\documentclass[]{aa}
\pdfoutput=1
\usepackage[varg]{txfonts}
\usepackage{xcolor}
\usepackage{placeins}
\usepackage[urlcolor=cyan, colorlinks=true, citecolor=blue, linkcolor=blue]{hyperref}
\bibpunct{(}{)}{;}{a}{}{,}

\title{The Cygnus Allscale Survey of Chemistry and Dynamical Environments: CASCADE. II. A detailed kinematic analysis of the DR21 Main outflow}
\author{I.~M.~Skretas\inst{1}, A.~Karska\inst{1,2}, F.~Wyrowski\inst{1}, K.~M.~Menten\inst{1}, H.~Beuther\inst{3}, A.~Ginsburg\inst{4}, A.~Hernández-Gómez\inst{5,1}, C.~Gieser\inst{6}, S.~Li\inst{3}, W.-J.~Kim\inst{7}, D.~A.~Semenov\inst{3}, L.~Bouscasse\inst{8}, I.~B.~Christensen\inst{1}, J.~M.~Winters\inst{8} \and A.~Hacar\inst{9} }

\institute{Max-Planck-Institut für Radioastronomie, Auf dem Hügel 69, 53121, Bonn, Germany 
\and Institute of Astronomy, Faculty of Physics, Astronomy and Informatics, Nicolaus Copernicus University, ul. Grudziądzka 5, 87-100 Toruń, Poland
\and Max Planck Insitute for Astronomy, Königstuhl 17, 69117 Heidelberg, Germany
\and Department of Astronomy, University of Florida, P.O. Box 112055, Gainesville, FL 32611-2055, USA
\and Tecnologico de Monterrey, Escuela de Ingenier\'{\i}a y Ciencias, Avenida Eugenio Garza Sada 2501, Monterrey, 64849, Mexico
\and Max-Planck-Institut für extraterrestrische Physik, Giessenbachstrasse 1, D-85748 Garching, Germany
\and I. Physikalisches Institut, Universität zu Köln, Zülpicher Str. 77, 50937 Köln, Germany
\and IRAM, 300 rue de la Piscine, Domaine Universitaire de Grenoble, 38406 St.-Martin-d'Hères, France
\and Department of Astrophysics, University of Vienna, Turkenschanzstrasse 17, 1180 Vienna, Austria
}

   \date{Received May 5, 2023; Accepted Sep 17, 2023}
	\titlerunning{A detailed kinematic analysis of the DR21 Main outflow}
	\authorrunning{I.~Skretas et al. 2023}

\abstract
{Molecular outflows are believed to be a key ingredient in the process of star formation. The molecular outflow associated with DR21 Main in Cygnus-X is one of the most extreme, in mass and size, molecular outflows in the Milky Way. The outflow is suggested to belong to a rare class of explosive outflows which are formed by the disintegration of protostellar systems.}
{We aim to explore the morphology, kinematics,
and energetics of the DR21 Main outflow, and compare those 
properties to confirmed explosive outflows to unravel the 
underlying driving mechanism behind DR21.}
{Line and continuum emission are studied at a wavelength of 3.6\,mm with IRAM 30 m and NOEMA telescopes as part of the Cygnus Allscale Survey of Chemistry and Dynamical Environments (CASCADE) program. The spectra include ($J= 1-0$) transitions of HCO$^+$, HCN, HNC, N$_2$H$^+$, H$_2$CO, CCH (among others) tracing different temperature and density regimes of the outflowing gas at high-velocity resolution ($\sim$ 0.8 km s$^{-1}$). The map encompasses the entire DR21 Main outflow and covers all spatial scales down to a resolution of ~3$\arcsec$ ($\sim$ 0.02 pc).}
{Integrated intensity maps of the HCO$^+$ emission reveal a strongly collimated bipolar outflow with significant overlap of the blue- and red-shifted emission. The opening angles of both outflow lobes decrease with velocity, from $\sim80$ to 20$^{\circ}$ for the velocity range from 5 to 45 km s$^{-1}$ relative to the source velocity. No evidence is found for the presence of elongated, $\lq\lq$filament-like$"$ structures expected in explosive outflows.
N$_2$H$^+$ emission near the western outflow lobe reveals the presence of a dense molecular structure which appears to be interacting with the DR21 Main outflow.}
{The overall morphology as well as the detailed kinematics of the DR21 Main outflow is more consistent with that of a typical bipolar outflow instead of an explosive counterpart.}
\begin{document}

\maketitle

\section{Introduction}

%General
Molecular outflows are a ubiquitous part of star formation arising from both high and low mass protostars \citep{Arce2007,Frank2014,bally2016}. A new type of outflows, 
formed by the disintegration of protostellar systems due to a merging event, has been proposed and tied to regions of high-mass star formation \citep{Bally2005,Zapata2009}. The massive outflow of DR21 Main is one of the proposed candidates for such explosive outflows \citep{Zap13}. The large angular extent of the DR21 Main outflow allows for a detailed analysis of its structure and properties, and their comparison to those of other explosive outflow candidates.

Molecular protostellar outflows range from highly collimated molecular jets like HH211 \citep{gueth1999} all the way to wide-angled outflows from high mass sources \citep{BeutherShepherd2005}. 
%KMM: MWC 349 is IT a protostar!
In general, outflows tend to appear \lq\lq narrower'' at higher velocities \citep{Bachiller1999}, and become less collimated as they evolve \citep{Beuther2005,Arce2006,Offner2011,Hsieh2023}.
They can vary significantly in size and energetics, with sizes from 0.1 pc up to pc scales and momentum rates between 10$^{-5}$ and 10$^{-2}$ M$_\odot$ km s$^{-1}$ yr$^{-1}$, for low mass sources and some O type stars, respectively \citep[e.g.,][]{Maud2015}. Some outflow properties, such as their mass, force and mechanical luminosity, correlate well with intrinsic parameters of their driving sources, e.g., the bolometric luminosity \citep{Bally1983,Cabrit1992,Wu2004} and the mass of the molecular gas envelope of their driving source \citep{Bontemps1996,Beuther2002}, suggesting a common driving mechanism in both the low and high mass sources. 
Clearly, molecular outflows play a critical role in regulating star formation by removing excess angular momentum and thus facilitating the further mass growth of a protostellar system \citep{Blandford1982,Machida2014}, and partly in dispersing the surrounding envelope, reducing the available mass reservoir \citep{Arce2006}. 

Due to their large size and the energy they carry, molecular outflows can have a significant impact to the surrounding interstellar medium (ISM) over different spatial scales. Firstly, at envelope scales (10$^3$-10$^4$ AU), powerful young outflows entrain and clear-out dense material giving rise to bipolar cavities \citep[e.g.][]{Gueth1997,Velusamy1998,Arce2004,Arce2005}. At core scales (0.1--0.3 pc), outflows are considered a significant contributor to the turbulence \citep{Myers1988,Zhang2005}. In addition, outflows from high mass young stellar objects (YSOs) might impact the morphology and even break apart the host molecular cloud \citep{Fuente1998,Benedettini2004}. 
Finally, the propagation of outflows through the surrounding dense material leads to the formation of shocks, which locally compress and heat the gas, and drive chemical processes enriching the ISM 
\citep[e.g.][]{Kau96,Flo10,Burk19}.

A newly proposed type of molecular outflows are the so--called  \lq\lq explosive dispersal outflows'', whose origin appears to linked to the disintegration of young stellar systems \citep{Bally2017,Rivera2021} or to protostellar mergers \citep{Bally2005}. The interpretation is limited due to the small sample of explosive-outflow candidates: Orion-KL \citep{Zapata2009}, DR21 Main \citep{Zap13}, G5.89 \citep{Zapata2019}, and IRAS 16076-5134 \citep{Ccolque2022}. 
Nevertheless, these explosive outflows share the following characteristics \citep{Zapata2009,Zap17}: (i) 
they consist of multiple straight, narrow and relatively isotropically distributed filament-like structures; (ii) these filament-like structures should all point towards the origin point of the explosive outflow and show an increase in velocity with the distance from the origin point akin to a Hubble flow; (iii) have a significant overlap of their blue- and red-shifted emission components. 
The filament-like structures of explosive outflows form because all material is simultaneously accelerated in the explosion. As a result, faster moving material has traveled further away from the source and is trailed by the slower parts of the outflow.
Overall, the properties of those outflows have been mostly studied using low$-J$ CO transitions at high angular resolution \citep[e.g.][]{Zapata2009,Zap13,Zapata2019}. The inner parts of the molecular outflows revealed multiple filament-like structures that make up the explosive outflows.  
At the same time, the lack of similar observations in other molecular tracers limits our understanding of their chemistry and various physical gas components.

The DR21 Main outflow is a particularly interesting explosive outflow candidate \citep{Zap13}, as it is one of the most massive ($M_\mathrm{out}>3000$ M$_\odot$) and energetic ($E_\text{kin} >$ 2 $\times 10^{48}$ erg) outflows detected in our Galaxy \citep{Garden1986,Gar91}, first in vibrationally excited $2.12~\mu$m line of shock-excited molecular hydrogen (H$_2$). DR21 Main itself is a compact HII region prominent at radio wavelengths. It is located in the Cygnus-X high-mass star-forming region/molecular cloud complex \citep{Leung1992}, at the southern end of the DR21 molecular ridge \citep{Dic78}, and at a distance of 1.5 kpc \citep{Rygl2012}.
The outflow appears bipolar, with the outflow lobes extending in a East--West direction 
\citep{Garden1986,Gar91h2,Gar92,Sch10}. High velocity low$-J$ CO emission has also been reported in the North--South direction \citep{Gar91}. The blue- and red-shifted parts of the outflow overlap significantly, suggesting that it extends very close to the plane of the sky \citep{Cru07}. It was initially suggested that the DR21 Main outflow is driven by a massive protostar, with $L_\mathrm{bol}$ of $\sim10^5-10^6$ $L_{\odot}$ \citep{Gar91,Gar92}, but such a source has not been yet identified \citep{Cru07}. 
The absence of a clearly detected driving source, along with the detection of some filament-like structures in CO (1 -- 0) emission, led \citet{Zap13} to suggest a possible explosive nature for the DR21 Main outflow.

In this work, we aim to study the morphology, kinematics and energetics of the DR21 Main outflow using observations in multiple molecular lines, sensitive to a range of physical conditions. We also aim to determine whether those characteristics of the DR21 Main outflow are consistent with those expected for explosive outflows or, rather, for typical protostellar outflows.

This work is a part of the Max Planck IRAM Observatory Program (MIOP) "Cygnus Allscale Survey of Chemistry and Dynamical Environments (CASCADE)" \citep{Beuther2022}. CASCADE aims to map significant parts of the Cygnus-X molecular cloud complex at high angular resolution and with a broad bandpass using the Northern Extended 
Array for Millimeter Astronomy (NOEMA) and the 30 m telescope, both
operated by the Institut de Radioastronomie Millim{\` e}trique (IRAM).
The combination of velocity-resolved single dish and interferometric observations offers the high resolution necessary to resolve the outflow structure without losing information on extended emission.
CASCADE aims to take advantage of these high quality observations to connect the transition of gas all the way from the large scales of molecular clouds down to the small scales of cores, to look for signs of collapse or feedback, to investigate the impact of star-forming cores to their surrounding, and to search for possible trends with evolutionary stage and more. The scope and goals of CASCADE are discussed in detail by \citet{Beuther2022}. 

% Table for observations
\begin{table*}[htb!]
\caption{Continuum and spectral line parameters for all lines covered by the CASCADE observations}
\label{table:observations} 
\centering 
\begin{tabular}{l c c c c c c} 
\hline\hline 
Species & Transition & \begin{tabular}[c]{@{}c@{}}Frequency\\  {[}GHz{]}\end{tabular}  & \begin{tabular}[c]{@{}c@{}}$\sigma_\text{rms}$\\  $\left[\frac{\rm \text{mJy}}{\rm \text{beam}}\right]$ \end{tabular} & \begin{tabular}[c]{@{}c@{}}Beam\\  {[}arcsec{]}\end{tabular} & \begin{tabular}[c]{@{}c@{}}E$_\mathrm{up}$\\  {[}K{]}\end{tabular} &  \begin{tabular}[c]{@{}c@{}}Log(A$_\mathrm{ij}$)\\  {[}s$^{-1}${]}\end{tabular} \\ 
\hline
Continuum & - & 82.028& 0.05& 2.80 $\times$ 2.54 & -& -\\
DCO$^+$& (1-0)  & 72.039& 15& 3.57 $\times$ 3.19 & 3.46& -4.16 \\
CCD & (1-0) & 72.108& 15& 3.57 $\times$ 3.19 & 3.46& -6.06 \\
DCN & (1-0) & 72.415& 16& 3.56 $\times$ 3.18 & 3.47& -4.88\\
SO$_2$ & (6$_{0,6}$-5$_{1,5}$) & 72.758& 14&3.54 $\times$ 3.16 & 19.15& -5.56 \\
HCCCN & (8-7) & 72.784& 14& 3.54 $\times$ 3.16 & 15.72& -4.53\\
H$_2$CO & (1$_{0,1}$-0$_{0,0}$) &72.838 & 16& 3.90 $\times$ 3.28 & 3.50& -5.09 \\
CH$_3$CN & (4$_k$-3$_k$) & 73.590& 12&3.51 $\times$ 3.13 & 8.83& -4.66 \\ 
DNC & (1-0) & 76.306& 10& 3.24 $\times$ 2.90 & 3.66& -4.79 \\
CH$_3$OH & (5$_{0,5}$-4$_{1,3}$)E &76.510 & 10& 3.24 $\times$ 2.89 & 47.93& -6.05 \\
NH$_2$D & (1$_{1,1}$-1$_{0,1}$) &85.926 & 10& 2.77 $\times$ 2.51 & 20.68& -5.71 \\
H$^{13}$CN & (1-0) &86.340 & 9& 2.75 $\times$ 2.50 & 4.14& -4.65 \\
H$^{13}$CO$^+$ & (1-0) & 86.754& 7 & 2.74 $\times$ 2.49 & 4.16& -4.41 \\
SiO & (2-1) & 86.847& 11 & 3.25 $\times$ 2.70 & 6.25& -4.53  \\
HN$^{13}$C & (1-0) &87.091 & 9 & 2.73 $\times$ 2.48& 4.18& -4.73 \\
CCH & (1-0) & 87.329& 10& 2.72 $\times$ 2.47 & 4.19& -5.90  \\
HNCO & (4$_{0,4}$-3$_{0,3}$) &87.925 &8 &2.71 $\times$ 2.45 & 10.55& -5.06  \\
HCN & (1-0) & 88.632& 13& 3.18$\times$ 2.64 & 4.25& -4.62 \\ 
HCO$^+$ & (1-0) & 89.189& 10&3.29 $\times$ 2.74 & 4.28& -4.38  \\
HNC & (1-0) & 90.664& 11& 3.12 $\times$ 2.57 & 4.35& -4.57 \\
HCCCN & (10-9) & 90.979& 9&2.64 $\times$ 2.39 & 24.01& -4.24  \\
CH$_3$CN & (5$_k$-4$_k$) &91.987 & 9& 2.61 $\times$ 2.36 & 13.24& -4.85 \\
H41$\alpha$& - & 92.034& 6& 2.60 $\times$ 2.36 & -& - \\
$^{13}$CS & (2-1) & 92.494& 10& 2.59 $\times$ 2.35 & 6.66& -4.85  \\
N$_2$H$^+$ & (1-0) & 93.174& 11& 2.58 $\times$ 2.33 & 4.47& -4.44 \\

\hline 
\end{tabular}
\tablefoot{
The rms noise level, $\sigma_\text{rms}$, is calculated for channels of 0.8 km s$^{-1}$ except for the case of H41$\alpha$ for which the channel width is 3.0 km s$^{-1}$. Rest frequencies are from NIST recommended rest frequencies \citep{Lovas2004} and the Cologne Database for Molecular Spectroscopy (CDMS, \citet{Muller2001}), while E$_\mathrm{up}$ and A$_\mathrm{ij}$ are taken from CDMS.
The conversion factor from Jy/beam to K, in the Rayleigh-Jeans limit, is $\sim$ 23.75 for a frequency of 89.189 GHz.}
\end{table*} 

% Figure of detections
\begin{figure*}
    \includegraphics[width=18cm]{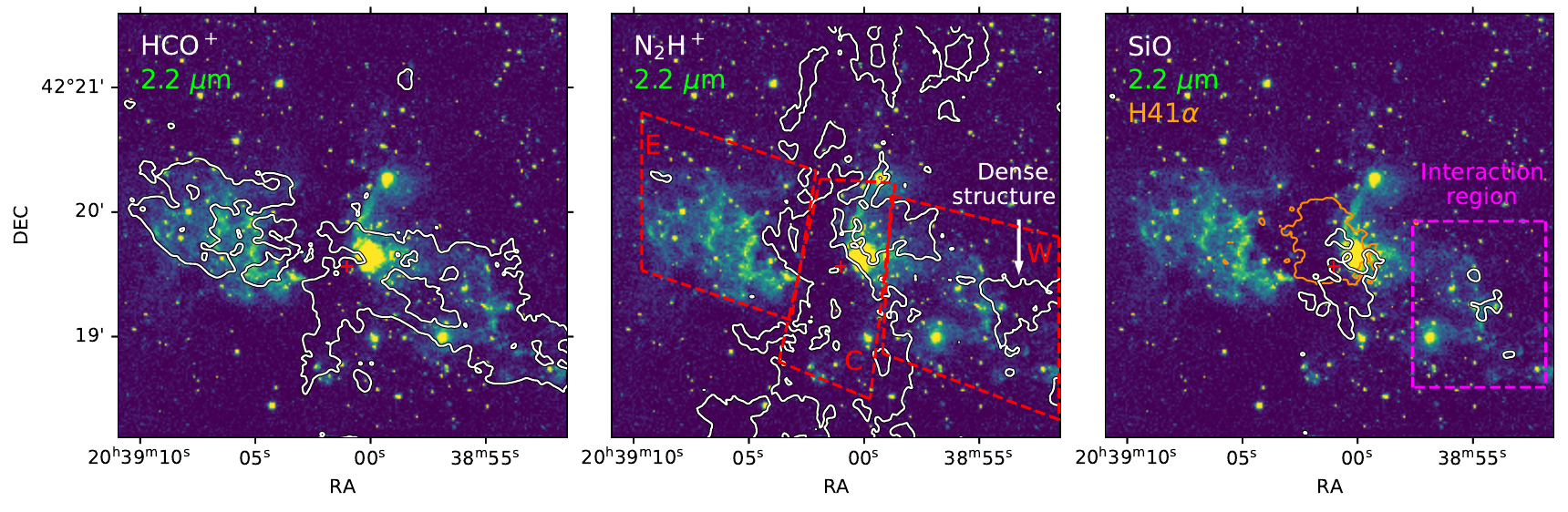}
    \caption{\textit{United Kingdom Infra-Red Telescope} (UKIRT) Wide Field Camera (WFCAM) continuum image of the DR21 Main region at 2.2 $\mu$m \citep{Warren2007} and the line emission in key gas tracers observed as part of CASCADE. Shock excited H$_2$ emission makes a significant contribution to the 2.2 $mu$m image,
    in particular to the lobes off the central region. White contours mark the 5$\sigma$ HCO$^+$ (left), N$_2$H$^+$ (middle) and SiO (right) emission.Intensities are integrated between -50 and 50, -20 and 10 ,and -20 and 20 km s$^{-1}$ for the HCO$^+$, N$_2$H$^+$ and SiO emisison respectively. The adopted origin point of the outflow is marked in all cases with a red cross. Red dashed lines (middle) mark the three separate areas of the DR21 Main region used to extract the spectra in Fig. \ref{fig:allspectra}, with \lq\lq E'' marking the eastern outflow lobe, \lq\lq C'' the central area and \lq\lq W'' the western outflow lobe. The magenta dashed box (right) marks the location of the interaction region shown in Fig. \ref{fig:interactioncontours}. Finally orange contours (right) show the 5$\sigma$ integrated intensity of H41$\alpha$ from -30 to 30 km s$^{-1}$.}
    \label{fig:h2images}
\end{figure*}
% -----------
% Table for detections 
\begin{table*}
\caption{Molecular line detections (at 5$\sigma$ level) in the area of the DR21 Main outflow} 
\label{table:detections} 
\centering 
\begin{tabular}{l l} 
\hline\hline 
Location & Species \\ 
\hline
Outflow (E -- W)& HCO$^+$, HCN\\
DR21 ridge - Dense gas (N -- S)& $^{13}$CS, CCH, H$_2$CO, H$^{13}$CO$^+$, HCCCN, HNC, N$_2$H$^+$, H$^{13}$CN, HN$^{13}$C\\
Sporadic or compact emission &CH$_3$CN\tablefootmark{a}, CH$_3$OH, DCN, DCO$^+$, DNC, H41$\alpha$, NH$_2$D, SiO\\
Non-detections & CCD, HNCO, SO$_2$ \\
\hline 
\end{tabular}
\tablefoot{
\tablefoottext{a}{Only a 3$\sigma$ detection.}
}
\end{table*} % ------- 

\begin{figure*}[h!]
\sidecaption
    \includegraphics[width=12cm]{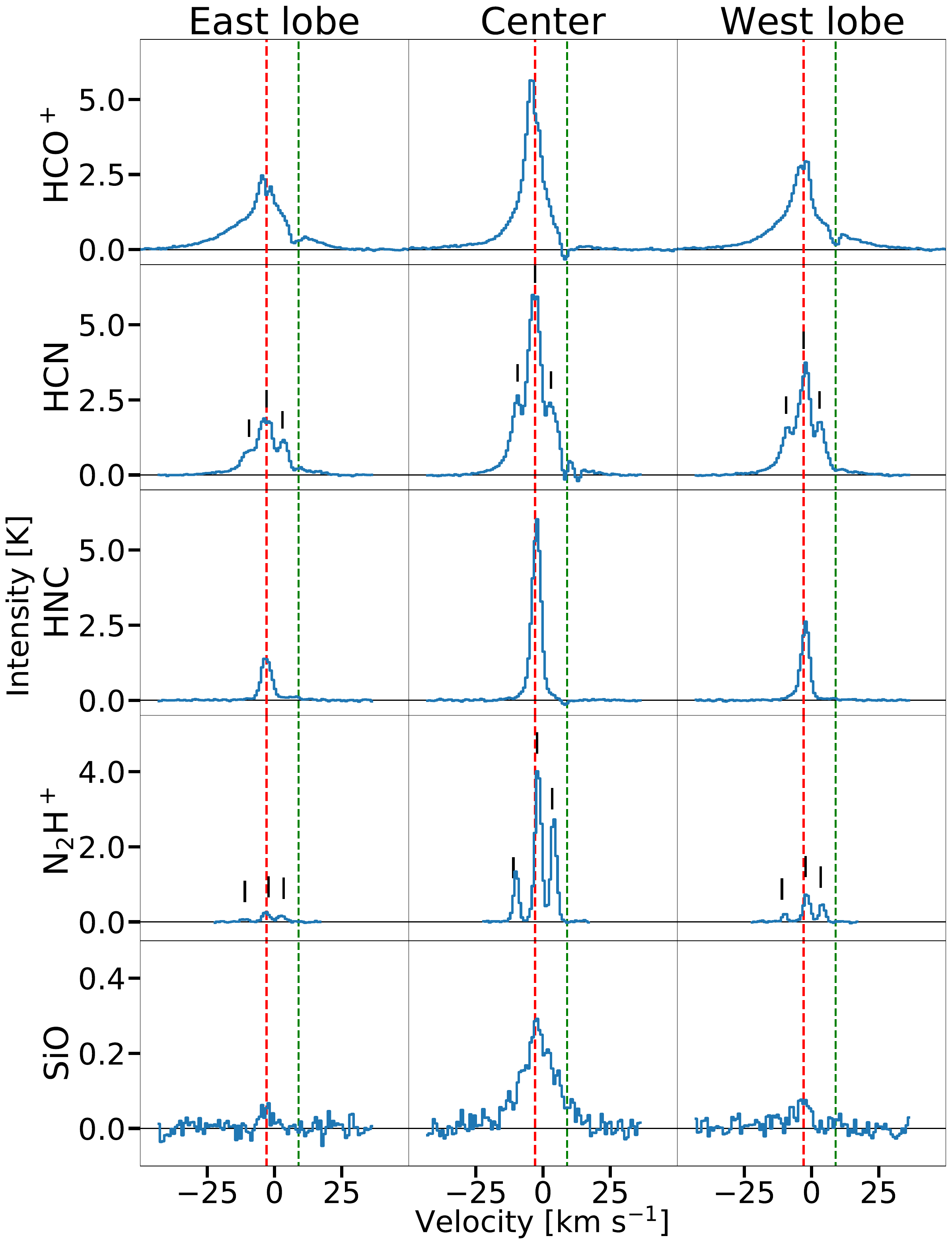}
    \caption{Averaged line profiles of the (1 -- 0) transitions of HCO$^+$, HCN, HNC, N$_2$H$^+$, and the (2 -- 1 ) transition of SiO toward the east lobe, center, and west lobe of the DR21 Main outflow (see Fig.~\ref{fig:h2images}). For HCN and N$_2$H$^+$, the velocities corresponding to hyperfine structure components are marked with black ticks. Red dashed lines show the source velocity, $-3$ km s$^{-1}$, green dashed lines mark the location of the absorption feature at 9 km s$^{-1}$ and the grey horizontal lines show the baselines. 
    }
    \label{fig:allspectra}
\end{figure*}

The paper is organized as follows. Section~\ref{sec:observations} describes the observations from CASCADE. Section~\ref{sec:results} presents line detections and maps of the DR21 Main outflow in several molecular transitions, and provides the analysis of outflow properties.
In Section~\ref{sec:discussion} the results are discussed and 
scenarios for the origin of the DR21 Main outflow are explored along with its interactions with the surrounding molecular cloud. Finally, Section~\ref{sec:summary} contains the summary and conclusions.

\section{Observations}
\label{sec:observations}

A detailed overview of the CASCADE program is given in \citet{Beuther2022}. In brief, CASCADE covers all high column density areas in the Cygnus-X molecular cloud complex using 40 mosaics, each covering 16 arcmin$^2$.
Each of the mosaics corresponds to 78 NOEMA pointings and was observed in both the C and D configurations. 
The observations have a total bandwidth of 16 GHz, 8 in each sideband, at the 3.6 mm window. The full bandwidth is covered with a spectral resolution of 2.0 MHz, but selected parts, surrounding the most important lines, are also covered by additional high resolution correlator units providing a spectral resolution of 62.5 kHz.   
The DR21 Main outflow is covered by two of the NOEMA mosaics, which were observed between 2020 May 29 and November 6. During that time, the array consisted of 10 antennas, yielding baselines between ~15 m and ~365 m. The strong quasars 3C345 and 3C273 were used as bandpass calibrators, MWC349 and 2010+723 were used for flux calibration and 2005+403, 2120+445, 2050+363 and 2013+370 were used for gain calibration.  
Complementary single-dish observations were carried out with the IRAM 30m telescope between 2020 February and July, in order to provide the missing short spacing information. These observations will be presented in detail in an upcoming paper by Christensen et al. (in prep.). 

The calibration and imaging of the data was done using CLIC and MAPPING software, which are part of the GILDAS package\footnote{\url{https://www.iram.fr/IRAMFR/GILDAS/}}). The NOEMA observations are combined with the IRAM 30m data using the UV\_SHORT task. The resulting single channel $\sigma_\text{rms}$ noise, for a channel width of 0.8 km s$^{-1}$, and beam sizes for all observed lines are summarized in Table \ref{table:observations}. 

% Velocity steps figure 
\begin{figure*} 
    \centering
    \includegraphics[width=18cm]{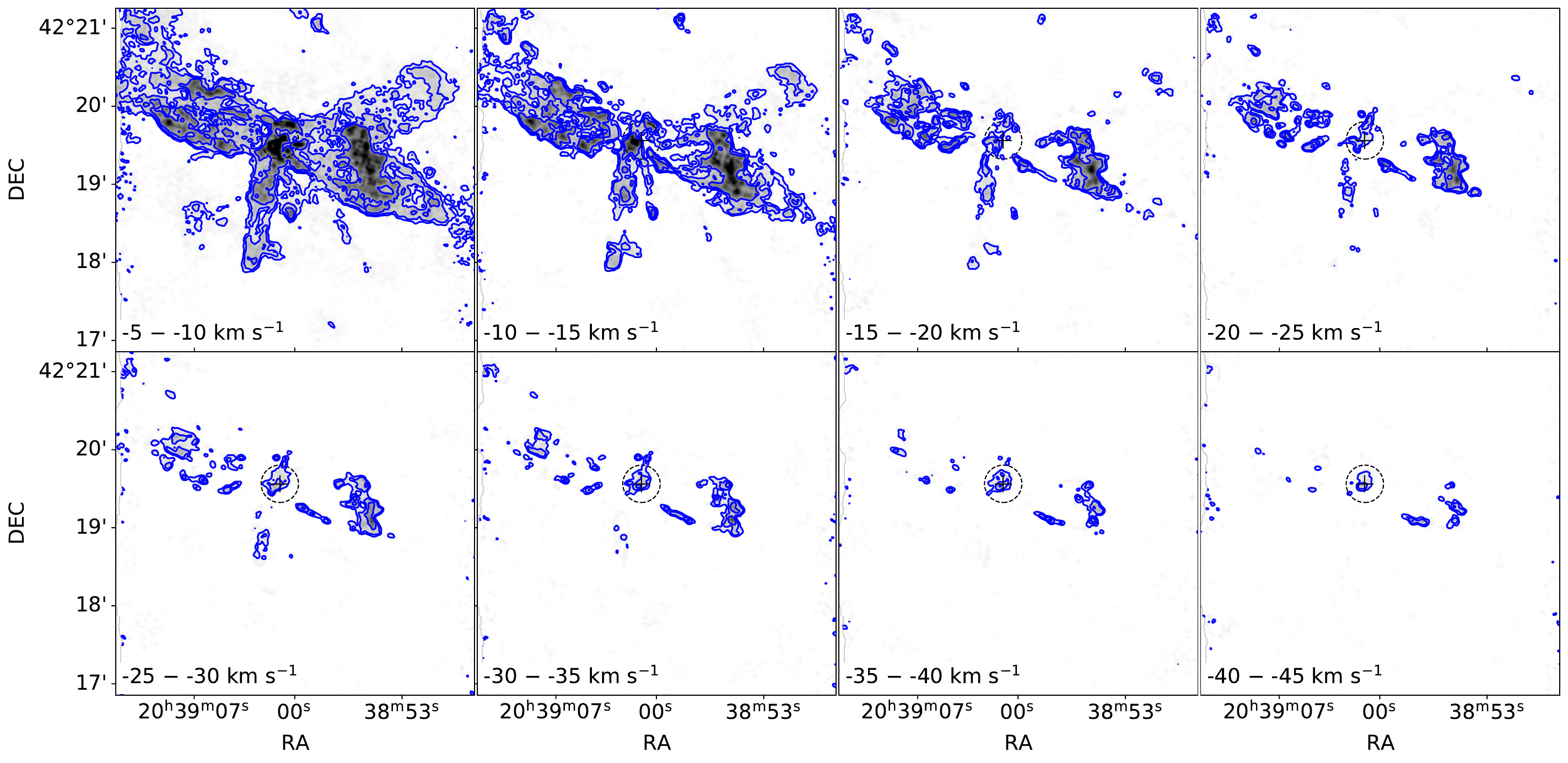}
    \includegraphics[width=18cm]{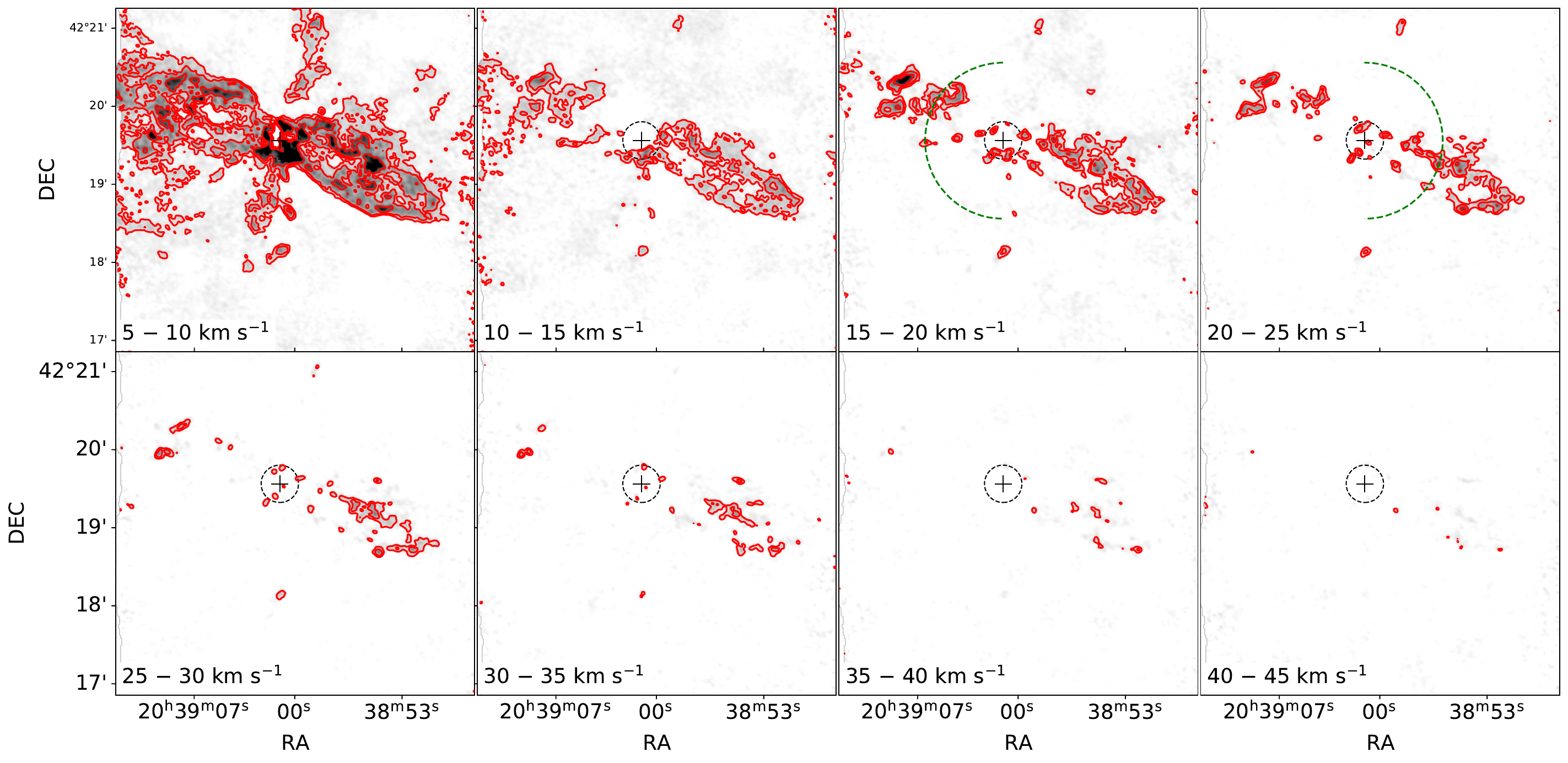}
    \caption{Channel maps of the DR21 Main outflow in HCO$^+$. Contours of the blue-shifted HCO$^+$ emission (in blue) and red-shifted emission (in red) are integrated over velocity steps of 5 km s$^{-1}$ and are plotted over the corresponding gray-scale. The full velocity range is from 5 to 45 km s$^{-1}$ relative to source velocity ($-$3 km s$^{-1}$) and the contour levels correspond to 5, 10 and 20 $\sigma_\text{rms}$. The black, dashed circle shows the area of the DR21--1 core in \citet{Cao19}. The green dashed lines denote the half-circles used to derive the outflow opening angles (see Fig. \ref{fig:opening angles}).}
    \label{fig:velsteps}
\end{figure*}
% --------
% Blue and red shifted integrated intensity
\begin{figure}
    \centering
    \includegraphics[width=9cm]{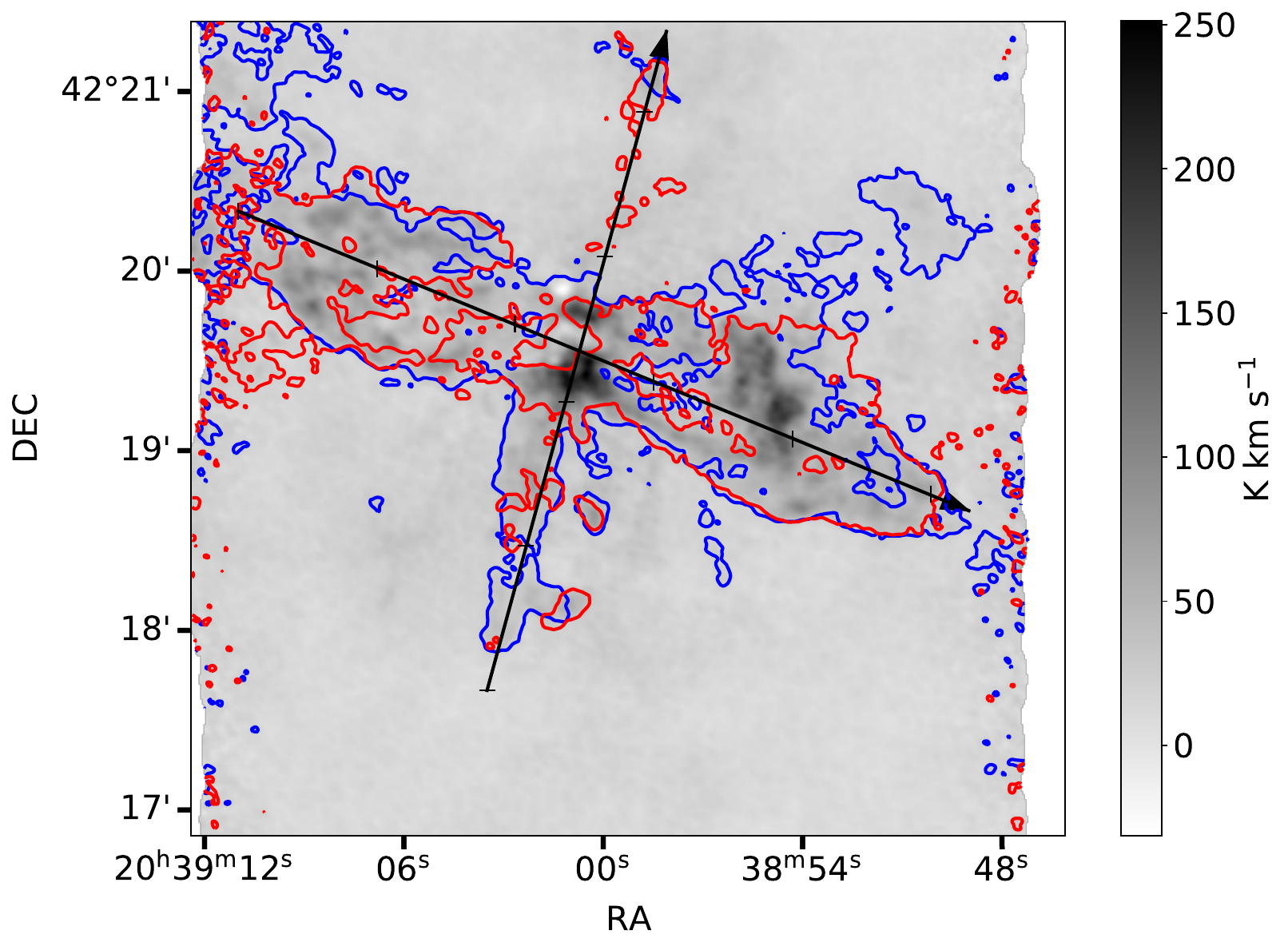}
    \caption{HCO$^+$ emission integrated from -50 to 50 km s$^{-1}$. Red and blue contours mark the redshifted (5 to 45 km s$^{-1}$) and blueshifted ($-$5 to $-$45 km s$^{-1}$) HCO$^+$ emission respectively. Contours correspond to 5 $\sigma_\text{rms}$ emission and the velocity ranges are given relative to the source velocity ($-$3 km s$^{-1}$). Black arrows mark the cuts for the PV diagrams (Figs. \ref{fig:PV_along} and \ref{fig:PV_across}) and the ticks mark distances of 50 arcseconds along the arrows.}
    \label{fig:HCO+_red_blue_contours}
\end{figure}
% -----------------

% PV diagrams --------------
\begin{figure}
    \centering
    \includegraphics[width=9cm]{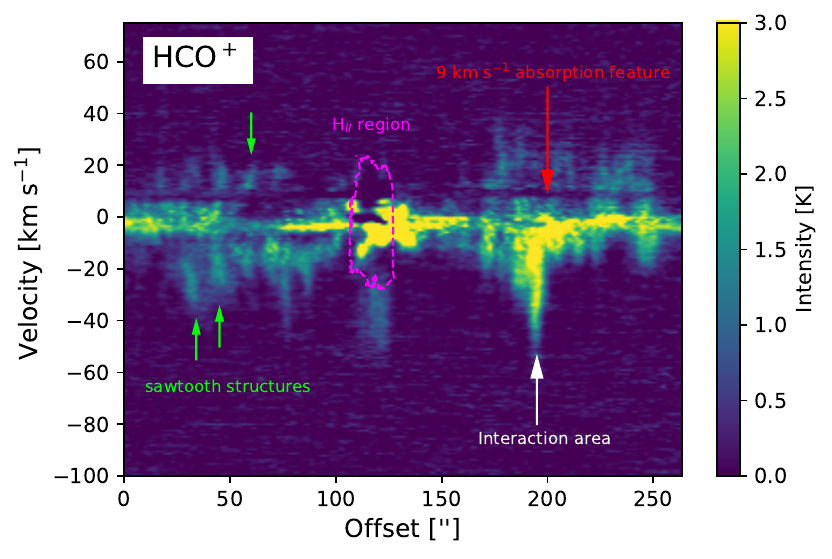}
    \caption{Position -- velocity diagram for HCO$^+$ emission along the DR21 Main outflow. The offset is measured from the edge of the eastern lobe towards the west. The arrow points to the 9 km s$^{-1}$ absorption feature \citep{Dic78}, the white arrow highlights the location of the interaction region (this work), the magenta dashed contour marks the H41$\alpha$ emission from the \ion{H}{ii} region, and the light green arrows mark examples of sawtooth pattern structures \citep{Sant2009}. }
    \label{fig:PV_along}
\end{figure}

\section{Results and analysis}
\label{sec:results}

We present the CASCADE observations for an area surrounding DR21 Main that covers the entirety of its outflow. Our data allow us to analyze the kinematics, morphology and energetics of the DR21 Main outflow and to contribute to a discussion about its nature.

\subsection{Molecular detections}
\label{sec:detections}
Several molecular lines are detected in the CASCADE observations of the DR21 Main (see Table \ref{table:detections}). The spatial distribution of the emission can be divided into three cases: (i) tracing the outflow (extended emission in the west--east (W--E) direction), (ii) tracing the DR21 ridge (extended emission in the north--south (N--S) direction) , and (iii) sporadic (compact emission that appears in multiple locations) or compact emission (see Fig. \ref{fig:h2images}). 
The different tracers can therefore be used to examine the morphology of the various gas components in DR21 Main. Integrated intensity contour maps for all the emission lines are shown in Appendix \ref{app:contours}.

The contour map of the HCO$^+$ integrated intensity (Fig. \ref{fig:h2images}, left panel) shows that most of the emission arises from the area of the outflow lobes and also appears to be in close agreement with the H$_2$ emission at 2.2 $\mu$m \citep[see also][]{Garden1986,Dav07}, associated with outflowing shocked gas. 
Interestingly, the HCO$^+$ (1 -- 0) contours reveal also the presence of hollowed out cavities in both outflow lobes lacking line emission, similar to early findings by \cite{Gar92}. The cavities are more prominent in the eastern outflow lobe, which appears entirely separated from the center of DR21 Main area. Therefore, while HCO$^+$ is well associated with the outflowing material, it traces most accurately the outer parts of the outflow cavities. In a similar fashion, line emission in HCN (1 -- 0) is also associated with the outflow, but shows a more compact pattern in the direction of the peaks of HCO$^+$ emission (see Fig. \ref{fig:appcontours1}).

In contrast, the N$_2$H$^+$ emission appears to trace the DR21 ridge along the N--S direction \citep[the middle panel of Fig. \ref{fig:h2images},][]{Wilson1990,Mot07}. This is expected as N$_2$H$^+$ is known to trace dense, cold, CO depleted gas \citep{Caselli2002,jorg04}. In addition to the ridge, N$_2$H$^+$ reveals also the presence of a molecular structure near the western lobe. This structure was previously detected in CS (2 -- 1) by \citet{Plambeck1990} who also reported the detection of a collisionally excited class I methanol maser in its interaction 
region with the outflow traced by H$_2$ emission.
Other molecules that are often associated with dense gas, like CCH (1 -- 0), HCCCN (both the (10 -- 9) and (8 -- 7) transitions), $^{13}$CS (1 -- 0) and HNC (1 -- 0) show distribution similar to that of N$_2$H$^+$ (1 -- 0) (see Figs. \ref{fig:appcontours1}, \ref{fig:appcontours2} and \ref{fig:appcontours3}).

Finally, SiO is detected close to the center of the outflow, but also shows very localized emission in an area of the western outflow lobe (Fig. \ref{fig:h2images}, right panel). Even though SiO typically traces shocks in the ISM \citep[e.g.][]{Martin1992,Schilke1997,Gusdorf2008a}, its emission peak in the center of DR21 might also originate from photo-evaporating ice mantles in the dusty envelope of a driving source(s) \citep[e.g.][]{Walmsley1999,Schilke2001}. On the other hand, the SiO emission detected near the western lobe could result from the interaction of the outflow and the the dense structure seen, for example, in N$_2$H$^+$, located there. This scenario is discussed in more detail in Sec \ref{sec:interaction}.

\begin{figure}
    \centering
    \includegraphics[width=9cm]{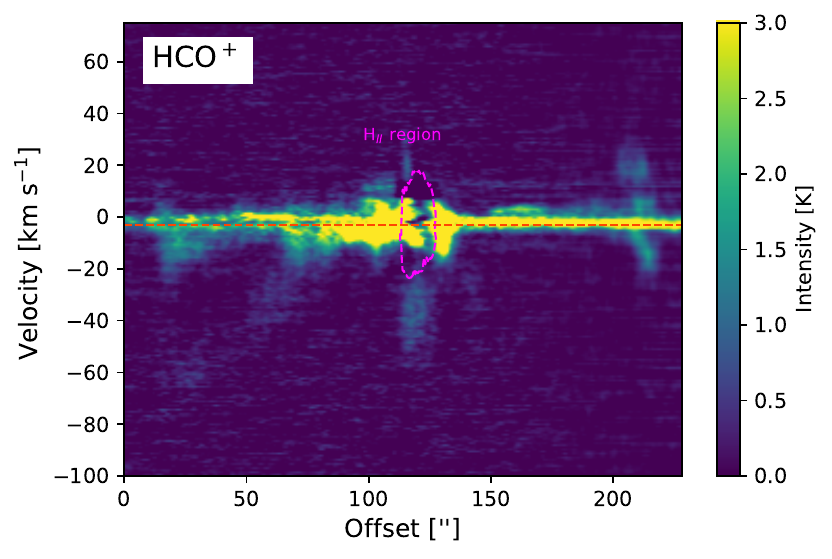}
    \includegraphics[width=9cm]{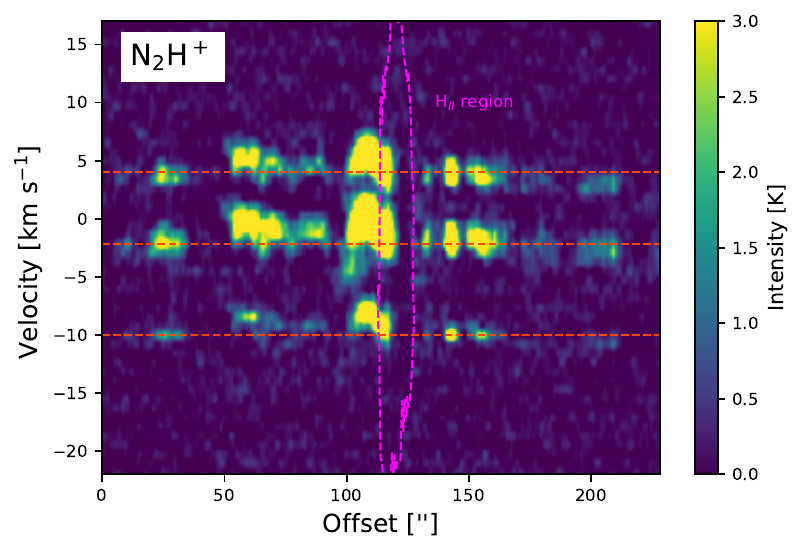}
    \caption{Position -- velocity diagrams for HCO$^+$ (top) and N$_2$H$^+$ (bottom) emission along the DR21 ridge. The offset is measured from South to North (see Fig. \ref{fig:HCO+_red_blue_contours} for the exact location of the cut). the magenta dashed contour marks the H41$\alpha$ emission from the \ion{H}{ii} region while the orange dashed lines mark the source velocity ($-$3 km s$^{-1}$). For the N$_2$H$^+$ line, 
    the velocities of the three resolved hyperfine structure components are marked at 4, $-$3 and $-$10 km s$^{-1}$, respectively.
    }
    \label{fig:PV_across}
\end{figure}

Figure \ref{fig:allspectra} shows the profiles of the strongest lines, averaged over the eastern outflow lobe, the central area of DR21 Main and the western outflow lobe (Fig. \ref{fig:h2images}). Similar spectra for the rest of the lines are presented in Appendix \ref{app:spectra}. 
The peak of the line emission lies at a velocity of $-$3 km s$^{-1}$, which corresponds to the velocity of the DR21 ridge \citep{Dic78}. The 9 km s$^{-1}$ feature, caused by more diffuse foreground material in the so-called ``extended W75 cloud'' \citep{Dic78,Nyman1983} can also be seen in some of the lines, but is most prominent in HCO$^+$ and HCN.
The profiles of both the HCO$^+$ and HCN lines display extended line wings and strong emission in both outflow lobes, thus confirming their association with the outflowing material.
In contrast, emission from molecules associated with the denser gas and the bulk of the DR21 ridge, like N$_2$H$^+$ or HNC have narrower emission lines and relatively weaker emission from the outflow lobes.

In summary, the CASCADE observations offer a clear view of the different gas components in the DR21 Main area. Most importantly, the HCO$^+$ is found to trace well the molecular outflow, N$_2$H$^+$ highlight the dense filament while SiO emission suggests the possibility of interaction between the outflow and the surrounding ISM, a scenario further explored in Section \ref{sec:interaction}. The release of a full line list, including unidentified lines for all targets of the CASCADE survey, will be presented in a future paper of the collaboration.

\subsection{The molecular outflow of DR21 Main}
\label{sec:dr21outflow}
The HCO$^+$ emission is one of the best tracers of the molecular outflow in DR21 Main and its distribution closely follows that of the H$_2$ emission (Section \ref{sec:detections}, Fig. \ref{fig:h2images}). Therefore, we use it to explore the kinematics as well as the outflow morphology at different velocities. In particular, we investigate the change of the opening angle with gas velocity.

Figure \ref{fig:velsteps} shows the spatial distribution of the HCO$^+$ (1 -- 0) emission integrated over velocity intervals of 5 km s$^{-1}$, in the range from 5 and 45 km s$^{-1}$ relative to source velocity ($v_\text{source} = -3 \text{ km s}^{-1}$) for the red-shifted emission, and from $-5$ to $-45$ km s$^{-1}$, for the blue-shifted emission. Most of the HCO$^+$ emission is elongated in the West--East direction, tracing the outflow lobes of a bipolar outflow (Fig. \ref{fig:h2images}). Some emission extends also in the North--South direction in a narrow range of velocities, suggesting that it is associated with the DR21 ridge. However, some of this emission  might also originate from the outflowing gas, as suggested by CO (1 -- 0) maps \citep{Gar91}.

Overall the velocity-channel maps (Fig. \ref{fig:velsteps}) show a rather symmetric morphology, but some small asymmetries can be noted.
Namely, the blue-shifted part of the outflow appears stronger and extends to higher velocities than its red-shifted counterpart. Similar behaviour is also seen between the two lobes of the outflow, with the western lobe appearing both brighter and having higher velocities that the eastern one. These small asymmetries are likely to arise due to the relative position of the outflow driving source compared to the bulk of material in the surrounding ISM.
In addition, the higher velocity HCO$^+$ emission seems to be detached from the origin point of the outflow \citep[see also, ][]{Gar92}. Due to the higher angular resolution of the current observations, we find cleared-out cavities in the outflow lobes.   
Finally, the known overlap of the blue- and red-shifted parts of the DR21 Main outflow, indicating that the outflow extends close to the plane of the sky, is clearly seen \citep[see Fig. \ref{fig:HCO+_red_blue_contours}, and ][]{Sch10}. We note, however, that we cannot estimate the inclination of the outflow more precisely because to the complexity of the ISM surrounding DR21 Main, e.g., the interaction region.

Figure \ref{fig:PV_along} shows the position--velocity diagram of HCO$^+$ along the DR21 Main outflow illustrating several key outflow structures. Firstly, the bright negative peak near the offset of 200$\arcsec$ indicates the location of the interaction region, where material is deflected into the line of sight. Secondly, an extended absorption feature is detected near the middle of the outflow, which corresponds to the \ion{H}{ii} region. Thirdly, the known absorption feature at 9 km s$^{-1}$, caused by more diffuse foreground material associated with W75, is also detected along the entire length of the outflow \citep{Dic78}. Finally, several structures are detected in the less disrupted Eastern lobe, which resemble a sawtooth pattern associated with the extremely high velocity component (EHV) of the low mass protostar IRAS 04166+2706 \citep{Sant2009}. A first estimate  using approximate values, derived from the PV diagram, for the maximum velocity of the structures ($\sim 20$ km s$^{-1}$) and the separation between them ($\sim 10\arcsec$) yields upper limits for the timescales between these knots of the order of $\sim 10^3$ yrs. This result corresponds to the upper limits of the timescales found between knots in the outflows of W43--MM1 \citep{Nony2020}.

Figure \ref{fig:PV_across} compares the velocities of HCO$^+$ with those of N$_2$H$^+$ in the direction of the DR21 ridge.  N$_2$H$^+$ is exclusively associated with the ridge (Fig. \ref{fig:h2images}) and shows 
three peaks corresponding to its hyperfine-splitted lines. All those components show a velocity gradient along the DR21 ridge, with velocities becoming increasingly blue-shifted North of DR21 Main. A similar gradient cannot be probed in the corresponding HCO$^+$ position--velocity diagram (Fig. \ref{fig:PV_across}) due to the complex line profiles. Similar to Figure \ref{fig:PV_along}, a strong central absorption feature exists, associated with the \ion{H}{ii} region. In addition, the extended emission in the North -- South is detected as a blue-shifted structure between offsets 50$\arcsec$ and 100$\arcsec$. 
 The lack of corresponding red-shifted emission favors the scenario that this emission is associated with the ridge and not an additional outflow, extending in the N--S direction. 

\begin{figure}
    \centering
    \includegraphics[width=8cm]{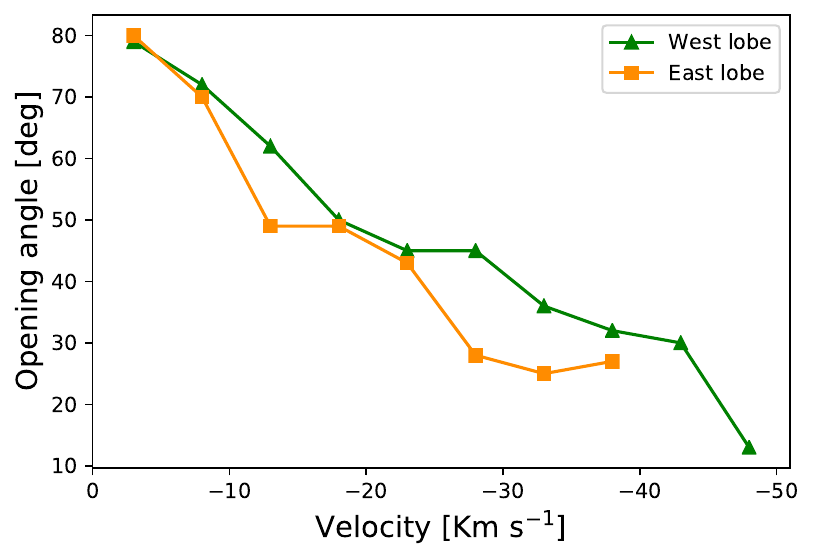}
    \includegraphics[width=8cm]{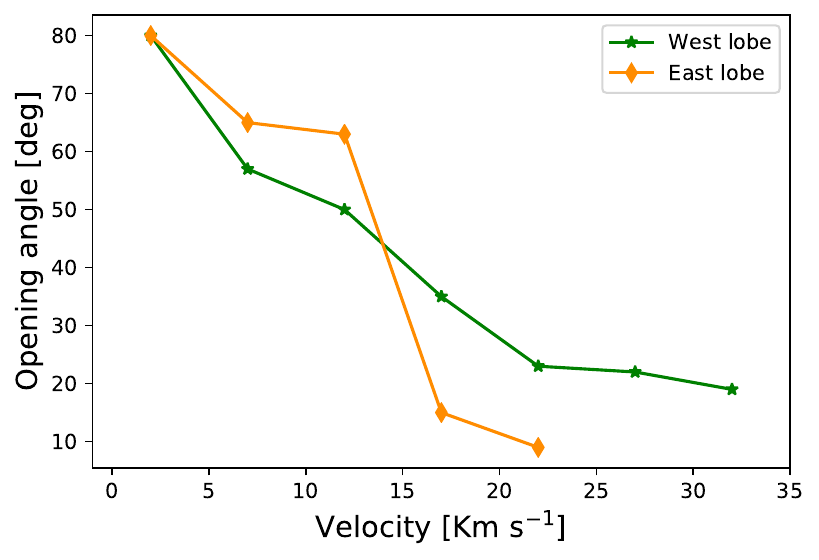}
    \caption{Opening angle of the DR21 Main outflow as a function of velocity using HCO$^+$ 1-0 line profiles. The angles are calculated every 5 km s$^{-1}$ from the source velocity up to $v_\mathrm{max} = 22, 32, -38$ \text{and} $-48$ km s$^{-1}$ for the east-red, west-red, east-blue and west-blue outflow lobe respectively. \textbf{Top}: Opening angles for blue-shifted outflow velocities. \textbf{Bottom}: Opening angles for the red-shifted outflow velocities.}
    \label{fig:opening angles}
\end{figure}

An important characteristic often used to describe outflows is their opening angle, a measure of how wide or collimated an outflow actually is. Here we calculated the opening angles for the DR21 Main outflow, separately for each lobe and for red- and blue-shifted emission, by examining the spectra of HCO$^+$ emission along a half-circle with radius approximately equal to half the extent of the corresponding outflow lobe (The exact location of these half-circles is shown in Fig. \ref{fig:velsteps}). The opening angle then corresponds to the angle between the location where the emission first becomes significant and the location where it drops further to noise level. The resulting opening angles for all different cases and for different velocities are plotted in Fig. \ref{fig:opening angles} over the corresponding velocity. Interestingly, the opening angles for all cases appear to be decreasing for higher velocities, a behavior that is expected in the case of a typical bipolar outflow, powered by a narrow and well collimated jet \citep[e.g.][]{Zhang2019,Raben2022}.

\subsection{Energetics of the outflow}

The spatially- and velocity-resolved observations of the DR21 Main outflow allow us to calculate key outflow properties such as, the outflow force, $F$, the rate at which the outflow injects momentum into its surrounding interstellar medium, the outflow mass, $M$, and its kinetic energy, $E_\text{kin}$. For all calculations, we use the HCO$^+$ (1 -- 0) emission found to trace well the molecular outflow (Section \ref{sec:detections}).

\begin{table}
\caption{Inclination correction factors used in the different methods of outflow force calculation}
\label{table:inclinations} 
\centering 
\begin{tabular}{c c c c c c} 
\hline\hline 
$i$\tablefootmark{a} ($\degr$) & 10 & 30 & 50 & 70 & Ref \\ 
\hline
$c_1$ & 0.28 & 0.45 & 0.45 & 1.1 & 1,2 \\ 
$c_2$ & 1.6 & 3.6 & 6.3 & 14 & 1 \\ 
$c_3$\tablefootmark{b} & 0.6 & 1.3 & 2.4 & 3.8 & 3 \\ 
\hline 
\end{tabular}
\tablebib{(1) \citet{Cabrit1990}; (2) \citet{Cabrit1992}; (3) \citet{Downes2007}}
\tablefoot{
\tablefoottext{a}{$i$ is measured from the line of sight.}
\tablefoottext{b}{Values are interpolated from Table 6 of \citet{Downes2007}}, where $\alpha = 90 - i$. 
}
\end{table}
\begin{table*}
\caption{Outflow parameters of the DR21 Main outflow} 
\label{table:energetics} 
\centering 
\begingroup
\renewcommand{\arraystretch}{1.5}
\begin{tabular}{r c c c c c c c c c} 
\hline \hline 
& \begin{tabular}[c]{@{}c@{}}$v_\text{max}$\\  {[}km s$^{-1}${]}\end{tabular} & \begin{tabular}[c]{@{}c@{}}$R_\text{lobe}$\\  {[}pc{]}\end{tabular}& \begin{tabular}[c]{@{}c@{}}$t_\text{dyn}$\\  {[}yr{]}\end{tabular} & \begin{tabular}[c]{@{}c@{}}$M$\\  {[}M$_\odot${]}\end{tabular} & \begin{tabular}[c]{@{}c@{}}$P$\\  {[}M$_\odot$ km s$^{-1}${]}\end{tabular} & \begin{tabular}[c]{@{}c@{}}$E_\text{kin}$\\  {[}erg{]}\end{tabular}& \begin{tabular} [c]{@{}c@{}}$ \dot{M}$\\  {[}M$_\odot$ yr$^{-1}${]}\end{tabular} & \begin{tabular}[c]{@{}c@{}}$F$\\  {[}M$_\odot$ km yr$^{-1}$ s$^{-1}${]}\end{tabular} & \begin{tabular}[c]{@{}c@{}}$L_\text{kin}$\\  {[}erg yr$^{-1}${]}\end{tabular}  \\ 
\hline
East lobe:~~~red & 38.6 & 0.74 & 4900 & 7 & 87 & 1.2 $\times$ 10$^{46}$ & 0.002 & 0.02 & 4.87 $\times$ 10$^{42}$ \\  
 blue &-62.2 & 0.75 & 3100 & 47 & 669 & 2.4 $\times$ 10$^{47}$ & 0.015 & 0.22 & 1.57 $\times$ 10$^{44}$ \\ 
\hline
West lobe:~~~red & 52.2 & 0.88 & 4300 & 13 & 135 & 1.8 $\times$ 10$^{46}$ & 0.003 & 0.03 & 8.14 $\times$ 10$^{42}$ \\ 
blue & -70.2 & 0.98 & 3600 & 57 & 1037 & 2.3 $\times$ 10$^{47}$ & 0.016 & 0.29 &1.26 $\times$ 10$^{44}$ \\ 
\hline
East+West:~~~red & -- & -- & -- & 20 & 222 & 3.0 $\times$ 10$^{46}$ & 0.005 & 0.05 & 1.30 $\times$ 10$^{43}$\\ 
 blue & -- & -- & -- & 104 & 1706 & 4.7 $\times$ 10$^{47}$ & 0.031 & 0.51 & 2.83 $\times$ 10$^{44}$ \\
\hline
Entire outflow & -- & -- & -- & 124& 1928 & 5.0 $\times$ 10$^{47}$& 0.036 & 0.56 & 2.96 $\times$ 10$^{44}$\\
\hline 
\end{tabular}
\endgroup

\end{table*}
To measure the force of the DR21 Main outflow, we use the so-called separation method introduced in \cite{vdm2013}, where the outflow force is calculated as:
\begin{equation}
    F_{\text{HCO}^+} = c_3 \times \frac{\displaystyle K \left( \sum_\text{j} \left[ \int_{v_\text{in}}^{v_\text{out,j}}T(v')v'\text{d}v' \right]_\text{j}\right) v_\text{max}}{\displaystyle R_\text{lobe}} .
    \label{eq:eq7}
\end{equation}
Here, $c_3$ is a correction factor for a given inclination angle of the outflow (Table \ref{table:inclinations}), $K$ is a conversion factor between the line integrated intensity and the molecular gas mass, the integral $\int_{v_\text{in}}^{v_\text{out,j}}T(v')v'\text{d}v'$ corresponds to the velocity weighted integrated intensity, $v_\text{max}$ is the maximum line-of-sight velocity in the outflow lobe, and $R_\text{lobe}$ is the length of the outflow lobe, while the sum runs over all pixels (j) that are part of the outflow. The conversion factor $K$ \citep[see  Appendix C of][]{vdm2013} is given by:
\begin{equation}
    K = \mu m_\text{H} A \frac{8 \pi k_\text{B}\nu^2}{h c^3 A_\text{ul}} \left[ \frac{\text{H}_2}{\text{HCO$^+$}}\right] \frac{Q{(T_\text{exc})}}{g_\text{u}}\text{e}^{E_\text{u}/T_\text{exc}}
    \label{eq:kappa}
\end{equation}
where $\mu$ is the mean molecular weight, $m_\text{H}$ is the hydrogen mass, $A$ is the observed area of the outflow, $[\text{H}_{2}/\text{HCO}^+]$ is the abundance ratio between H$_2$ and HCO$^+$, $Q{(T_\text{exc})}$ is the partition function at a specific excitation temperature $T_\text{exc}$, $g_\text{u}$ is the degeneracy of the upper level of the observed transition, $E_\text{u}$ is the upper level energy in Kelvins, $\nu$ is the frequency of the observed transition in Hz, $c$ is the speed of light, $k_\text{B}$ is Boltzmann's constant, $h$ is Planck's constant and $A_\text{ul}$ is Einstein A coefficient for the transition in s$^{-1}$.  We assume a single excitation temperature of 40 K \citep{Gar91}. The abundance ratio of H$_2$ over HCO$^+$ in high-mass star-forming regions has been found to range between 2 $\times$ 10$^9$ down to 3 $\times$ 10$^7$ \citep{Godard2010,Gerner2014}. In this work, we adopt an abundance ratio of H$_2$ over HCO$^+$ of $1.6\times10^{8}$ \citep{Gar92}, which is well within the above range and was estimated for DR21 Main. The corresponding value of the partition function and the remaining molecular data are taken from Splatalogue\footnote{\url{https://splatalogue.online/advanced1.php}} using the CDMS catalogue \citep{Muller2001}.

The inner velocities for the integration ($v_\text{in}$) are $-$5 and 5 km s$^{-1}$ relative to the source velocity for the blue- and red-shifted parts of the emission, excluding the innermost 10 km s$^{-1}$ in order to avoid contamination from the cloud material. 

\begin{figure}[h!]
    \centering
    \includegraphics[width=\columnwidth]{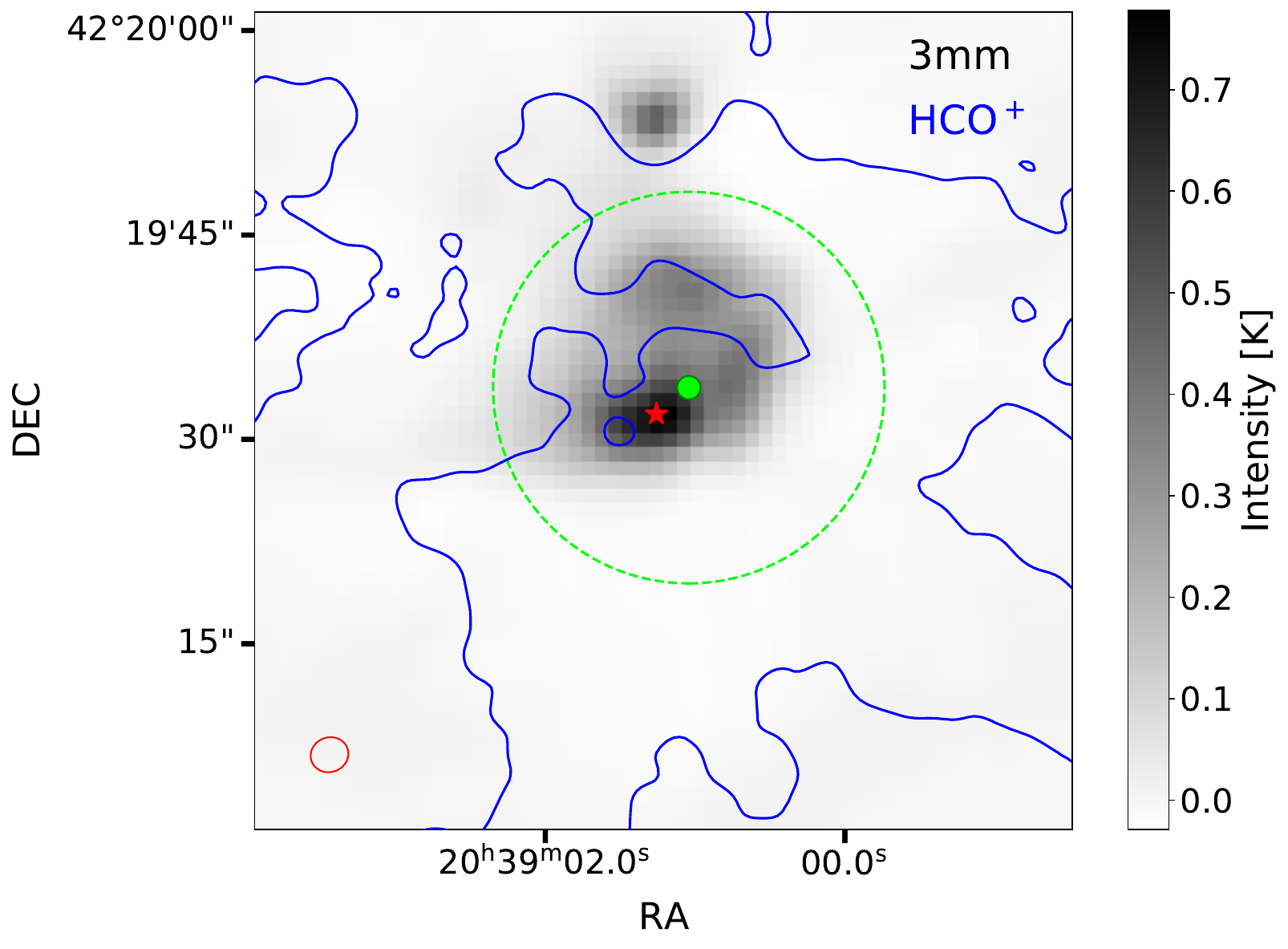}
    \caption{3mm continuum emission in the area of DR21 Main. The peak of the continuum emission is marked with a red star, while the green circle marks the location, and the green dashed line the FWHM, of the dense core DR21-1 \citep{Cao19}. Blue contours mark the integrated HCO$^+$ intensity (-50 to 50 km s$^{-1}$), while the beam of the continuum observations is noted with the red ellipse.}
    \label{fig:continuum}
\end{figure}

The calculation of the length of the outflow lobe, $R_\mathrm{lobe}$, requires information about the origin point of the outflow. Here, we adopt it as the position of the dense core DR21-1 from \cite{Cao19} \citep[see also core N46 in][]{Mot07}, which is located close to the center of the two outflow lobes and at the DR21 ridge. The location of the core agrees also well with the peak of the 3mm continuum emission, as shown in Fig. \ref{fig:continuum}. Higher resolution observations of this area would be required to determine the exact location and nature of the driving source, which is outside of the scope of this paper. 

% Looking for filaments figures ----- 
\begin{figure*}[ht!]
    \centering
    \includegraphics[width=2\columnwidth]{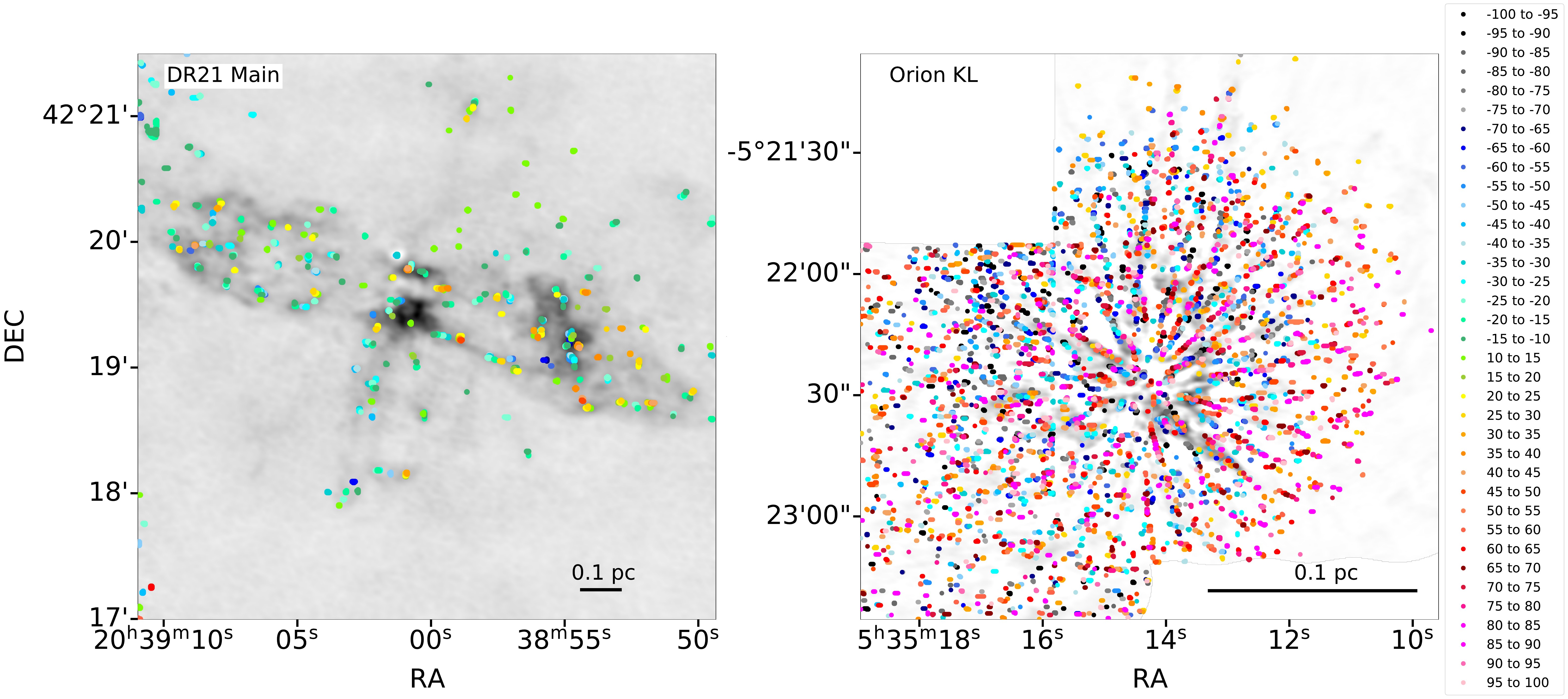}
    \caption{Distribution of gas velocities associated with the outflows in DR21 (left) and Orion KL \citep[right, ][]{Bally2017}. Each point shows a local intensity peak of the HCO$^+$ emission (above 2 $\sigma$) integrated in velocity steps of 5 km s$^{-1}$. The color of the points signifies the corresponding velocity steps and are shared for the two plots. Background grey-scale shows the integrated HCO$^+$ (left panel) and CO emission (right panel) over the entire velocity range of the outflows (from $-$70 to 70  km s$^{-1}$ for  DR21 Main and $-$100 to 100 km s$^{-1}$ for Orion KL).}
    \label{fig:filaments}
\end{figure*}
 
The gas mass carried by the outflow is obtained from:
\begin{equation}
    M = K \left( \sum_\text{j} \left[ \int_{v_\text{in}}^{v_\text{out,j}}T(v')\text{d}v'\right]_\text{j}\right). 
\end{equation}
Subsequently, the time-averaged kinetic energy of the outflow is calculated as:
\begin{equation}
E_\text{kin} = \frac{1}{2}M \langle v \rangle^2 ,
\end{equation}
and its momentum as:
\begin{equation}
    P = \displaystyle K \left( \sum_\text{j} \left[ \int_{v_\text{in}}^{v_\text{out,j}}T(v')v'\text{d}v' \right]_\text{j}\right).
    \label{eq:momentum}
\end{equation}

The dynamical time, which is an estimate of the lifetime of the outflow, is then measured from:
\begin{equation}
    t_\text{dyn} = \frac{R_\text{lobe}}{v_\text{max}}
\end{equation}
This, in turn, allows the calculation of the mass loss rate:
\begin{equation}
    \dot{M} = \frac{M}{t_\text{dyn}},
\end{equation}
and the power of the outflow:
\begin{equation}
    L_\text{kin} = \frac{E_\text{kin}}{t_\text{dyn}}.
\end{equation} 
The calculations are performed for the east and west outflow lobes, and for the red- and blue-shifted emission, separately. The resulting outflow properties are presented in Table \ref{table:energetics}. 

The outflow mass and kinetic energy can be compared with the results from \cite{Gar92}, where observations of the transition of HCO$^+$ (1 -- 0) were analyzed. Here, we use the high velocity component from \citet{Gar91} and scale it to the same distance of DR21, as adopted in this work. The outflow mass for the high-velocity component, of $\sim$120 M$_{\odot}$ (see Table \ref{table:energetics}) is a factor of 2-5 higher than the corresponding outflow mass in \cite{Gar92}. The outflow extent and the area covered by the observations are similar in both studies; \citet{Gar91} obtain the $R_\mathrm{lobe}$ of $\sim$1.7 pc. 
Thus, the difference is likely due to the velocity limits adopted in \cite{Gar92}, which exclude a significant part of gas mass at velocities close to the source velocity ($v_\mathrm{in}$ from $-$12.5 to $-$42.5 km s$^{-1}$). In fact, the outflow kinetic energy, which accounts for the relevant range of velocities, is fully consistent: $E_\text{kin}$ of 5.0$\times10^{47}$ erg (Table \ref{table:energetics}) 
is within the range of $\sim$2.5--5.0 $\times10^{47}$ erg reported in \cite{Gar92} using HCO$^+$ and about a factor of 4 lower than the total energy measured using CO \citep{Gar91}. For a more thorough discussion of the DR21 Main outflow properties, with respect to both low- and high-mass protostars, see Section \ref{sec:forcemass}.

The calculation of the outflow parameters includes a few assumptions that need to be addressed. Firstly, the conversion factor $K$ is accurate only when the observed emission is optically thin. In the case of DR21 Main, the  HCO$^+$ emission is optically thin in the outflow lobes, but not in the central area \citep[][]{Gar92}. Using our H$^{13}$CO$^+$ observations, which appear to trace the dense ridge (Section \ref{sec:results}), we estimate a $\tau$ of $12.5$ for the DR21 Main center. For that, the central region is excluded from the calculation of the outflow parameters. 
Similarly, gas with velocities within $\sim$5 km s$^{-1}$, from the velocity of the N--S filament is also excluded. 
Noteworthy, even though a significant part of the outflow material is often found at low velocities, its impact on $F$ or $E_\text{kin}$ is not as significant due to the dependence of those parameters on $v^2$. 
Secondly, the correction factor of inclination angle, $c_3$, is available only for the inclination angles of 10$\degr$, 30$\degr$, 50$\degr$, and 70$\degr$ (Table \ref{table:inclinations}). In this work, $c_3$ of 3.8 is assumed (corresponding to 70$\degr$), but since the outflow of DR21 Main appear to be close to 90$\degr$ (see Section \ref{sec:dr21outflow}), the correction is most likely underestimated by a factor of $\lesssim2$. Finally, we assumed a uniform excitation temperature of the gas along the outflow, which likely differs by a factor of a few depending on the position of line emission. The increase of excitation temperature from 40 to 80 K would lead to the increase of the $K$ parameter by a factor of $\sim 1.9$. Thus, the variations of $T_\text{ex}$ along the outflow are not expected to significantly impact the results.

To summarize, the parameters of the DR21 Main outflow, calculated using the 
HCO$^+$, provide a useful diagnostic of outflow energetics. The calculations are consistent with previous work by \cite{Gar92} using the same tracer and transition. Due to the optical thickness of the emission and the adopted velocity limits, the mass of the outflow as well as the related parameters are lower limits to the actual parameters.

\section{Discussion}
\label{sec:discussion}

\subsection{The nature of the DR21 Main outflow}
\label{sec:dr21nature} 
The nature of the DR21 Main outflow is still a topic of discussion since it has been proposed to belong to the class of explosive outflows \citep{Zap13}. The detailed analysis of the morphology and the kinematics of the outflow using HCO$^+$ observations reveals a rather well defined bipolar outflow structure reminiscent of that of a typical protostellar outflow (Section \ref{sec:dr21outflow}). The strong overlap of red- and blue-shifted HCO$^+$ emission is indeed a property attributed to explosive outflows \citep{Zap17}, however it can also appear in the case of a bipolar outflow that extends along the plane of the sky with the red- and blue-shifted emission arising from the sideways expansion of the outflow lobes. Additionally, the apparent decrease of the DR21 Main outflow's opening angle with increasing velocities (see Fig. \ref{fig:opening angles}) is characteristic of bipolar outflows that are powered by a narrow collimated jet but seems unlikely to happen in the case of explosive outflows.  

Figure \ref{fig:filaments} shows a side-by-side comparison of local intensity peaks at multiple velocity steps between the DR21 Main outflow and the Orion KL outflow. For the DR21 Main outflow the HCO$^+$ data presented in this work have been used, while for Orion archival ALMA CO (2 -- 1) data were used (Project ID: 2013.1.00546.S, PI: John Bally, \citet{Bally2017}). 
The emission in Orion (right) can be seen to consist of multiple, well defined filament-like structures which also display clear velocity gradients along their length with higher absolute velocities being further away from the outflows point of origin. This behaviour is probably the most distinctive characteristic of explosive outflows and is absent in the case of the DR21 Main outflow (left). More precisely, in DR21 Main there are very few distinct filament-like structures which also do not show any significant velocity gradients along their length. Moreover, the structures that do exist appear to be tracing the cavity walls of the outflow lobes contrary to the more random distribution that the filament-like structures have in Orion. We note though that DR21 is located significantly further, at 1.5 kpc \citep{Rygl2012}, than Orion KL at $\sim$400pc \citep{Menten2007,Kounkel2017}, which means that the physical scales of the two outflows in Fig. \ref{fig:filaments} are significantly different.

Taking all the above into consideration, it appears that the outflow of DR21 Main resembles more a typical bipolar outflow that is driven by a protostar rather than an explosive outflow. However, the observations presented here do not reveal a single, compact emission that could be associated with the driving source (Appendix A). Further observations are required in order to discern the protostar or protostellar system behind such an exceptionally powerful bipolar outflow.

\subsection{Outflow energetics}
\label{sec:forcemass} 

% Outflow energetic correlation plots
\begin{figure*}[ht!]
  \begin{minipage}[t]{.5\textwidth}
  \begin{center}  
      \includegraphics[angle=0,height=7.6cm]{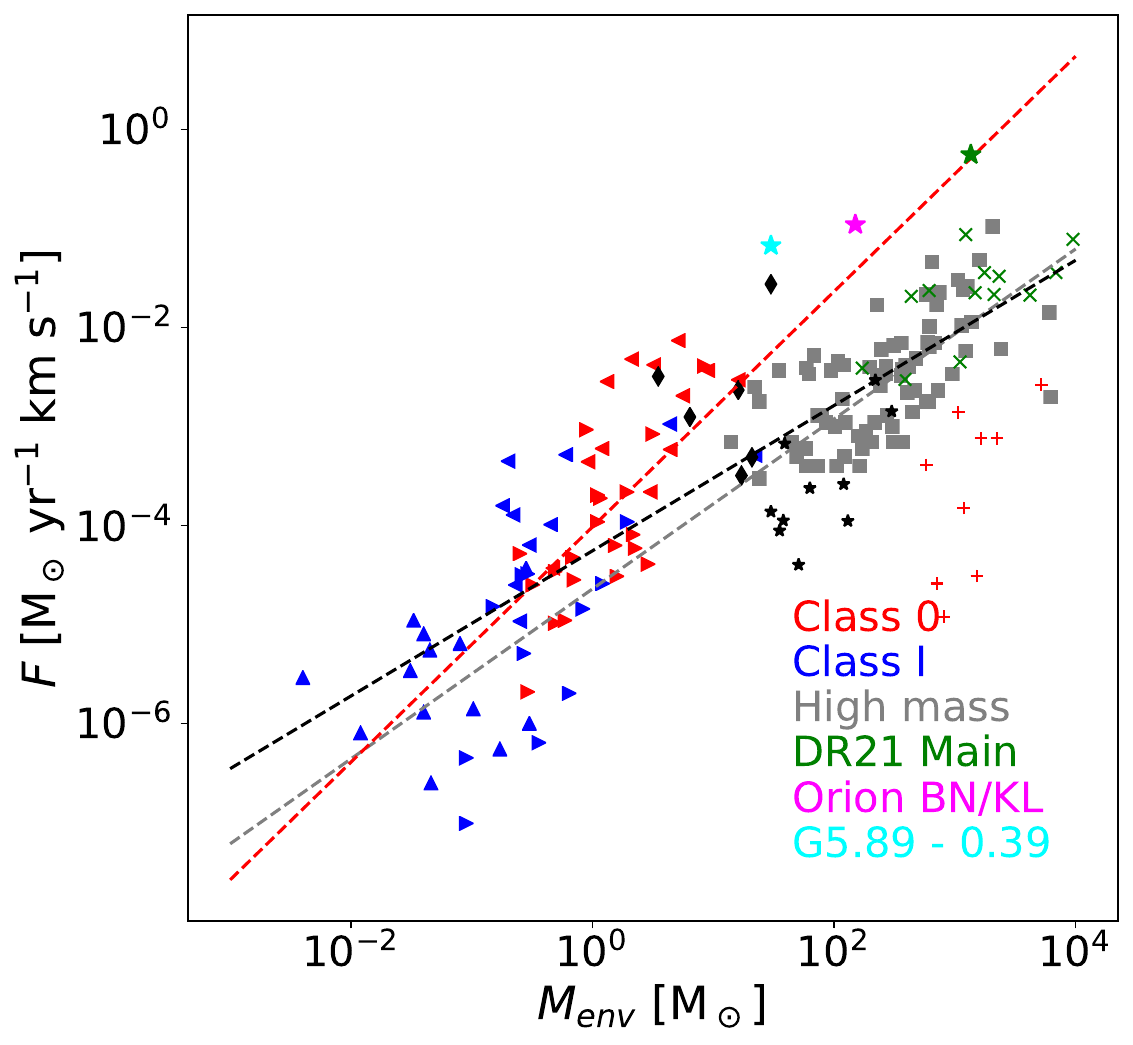} 
    \end{center}
  \end{minipage}
  \hfill
  \begin{minipage}[t]{.5\textwidth}
  \begin{center}         
    \includegraphics[angle=0,height=7.6cm]{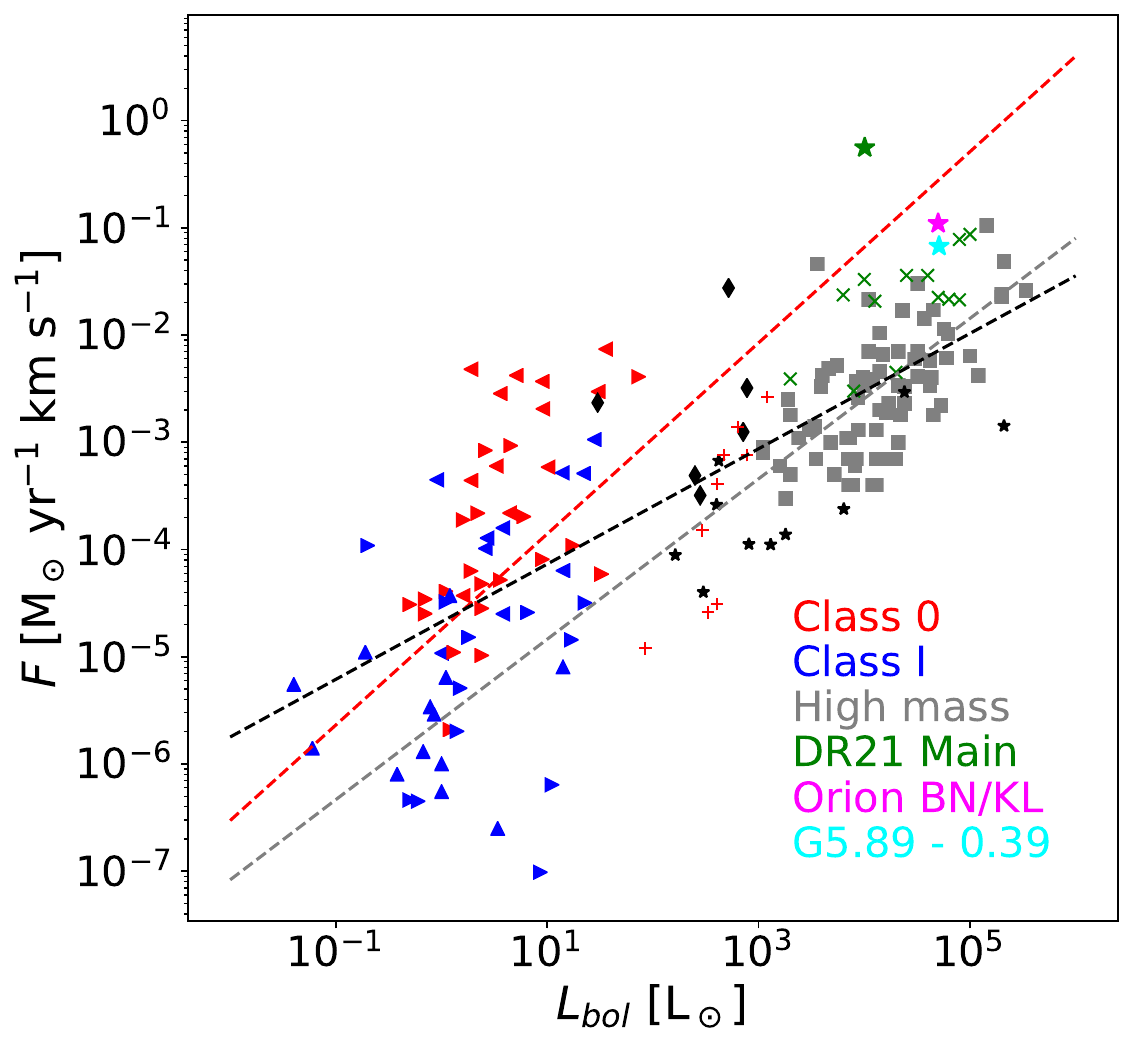} 
    \end{center}
  \end{minipage}
    \hfill
        \vspace{-0.2cm}
        \caption{\label{fig:forcemass} Outflow force over the envelope mass of the driving source for various protostellar sources. Right facing triangles represent low-mass sources from \citet{Mottram2017}, left facing triangles mark sources taken from \citet{Yildiz2015} and upwards are from \citet{vdm2013} while in blue are the Class I sources and in red the Class 0. Black diamonds mark intermediate mass sources \citep{vankempen2009}, grey squares mark high mass sources from \citet{Maud2015}, green \lq\lq$\times$'' mark high mass sources from \citet{Beuther2002} and black stars mark high mass sources in Cygnus \citep{Skretas2022}. The red crosses mark a sample of high-mass 70$\mu$m dark sources \citep{Shanghuo2020}. The Cyan star marks G5.89-0.39, the magenta one marks Orion KL and the green one represents the DR21 Main outflow. The dashed, black line shows the best fit to the outflow force - envelope mass correlation for all sources, while red and grey show the best fits for the low- and high-mass sources respectively.}
\end{figure*}
Multiple correlations have been reported connecting the energetics of protostellar outflows with the properties of their driving sources. For example, the correlation of the outflow force with the envelope mass and bolometric luminosity are attributed to a connection between mass accretion rates and outflow activity \citep{Beuther2002,duarte2013,Mottram2017}. The existence of such correlations allows for a  direct comparison of the properties of DR21 Main outflow and those of other protostellar and explosive outflows. 

Figure \ref{fig:forcemass} shows the comparison of the  outflow force of the DR21 Main outflow and those of low-, intermediate- and high-mass protostars \citep{Beuther2002,vankempen2009,vdm2013,Yildiz2015,Maud2015,Mottram2017,Shanghuo2020,Skretas2022}. We calculate the Pearson coefficients and the corresponding significance ($\sigma$) for both correlations \citep[see e.g., ][]{Mar10}. The outflow forces of high-mass protostars are found to correlate strongly both with their envelope masses (6.1$\sigma$) and bolometric luminosities (6.5$\sigma$); the correlation extends also to lower masses. The source sample from \citealt{Shanghuo2020} represents sources at the very early stages of their evolution (70 $\mu$m dark clumps), and are therefore expected to have a low $L_\text{bol}/M_\text{env}$ ratio. That is why they display relatively large envelope masses with respect to other sources of similar energetic parameters.

The outflow force of DR21 is higher than those of other high-mass protostars, including the other two explosive-outflow candidates: G5.89-0.39 \citep[$L_\mathrm{bol}$ of 4.1$\times$ $10^4$ L$_{\odot}$, $M_\mathrm{env}$ of 140 M$_{\odot}$, ][]{vT2013,karska14a} and Orion BN/KL \citep[$L_\mathrm{bol}$ of 5$\times$ $10^4$ L$_{\odot}$, $M_\mathrm{env}$ of 150 M$_{\odot}$,][]{Downes1981,Genzel1989}. For DR21, we adopt $L_\mathrm{bol}$ of 1.0$\times$ $10^4$ L$_{\odot}$ and $M_\mathrm{env}$ of 1355 M$_{\odot}$, which account for the new distance to the source and the peak of the Spectral Energy Distribution \citep{Cao19}.

DR21 shows also enhanced values of the mass outflow rate and outflow kinetic luminosity, but falls within the range of other high-mass protostars when the outflow mass, power, and kinetic energy are concerned (Appendix C). Those differences are therefore the largest for parameters involving the outflow dynamical time, and as such are related with $v_\mathrm{max}$ which in turn depends on the inclination angle. However, the inclination of DR21 on the sky could only introduce a factor of $\sim$4 difference in the derived parameters -- much less than the enhancement in outflow properties with respect to typical high-mass protostars. 

Orion BN/KL and G5.89-0.39 also show relatively high mass outflow rates, and kinetic energy and luminosity, which might suggest this could be a common characteristic of explosive-outflow candidates. Assuming that the underlying physical mechanism for explosive outflows is different than for typical outflows,
there is presently no theoretical expectation that explosive outflows should follow the $F$-$M_\mathrm{env}$ and $F$-$L_\mathrm{bol}$ correlations (Figure \ref{fig:forcemass}). However, the sample of these objects is too small to be conclusive.

The DR21 Main outflow is found to be a bipolar outflow (Section \ref{sec:dr21nature}). Therefore, the enhanced outflow force of DR21 might indicate the presence of scatter for the high-mass sources, similar to the one measured in the outflow properties of low-mass YSOs (Figure \ref{fig:forcemass}). In any case, the high outflow force of DR21 Main outflow is consistent with it being one of the most powerful outflows in the Galaxy.

\subsection{Interaction at the western outflow lobe}
\label{sec:interaction}

% Interaction area contours  
\begin{figure}
    \centering
    \includegraphics[width=8cm]{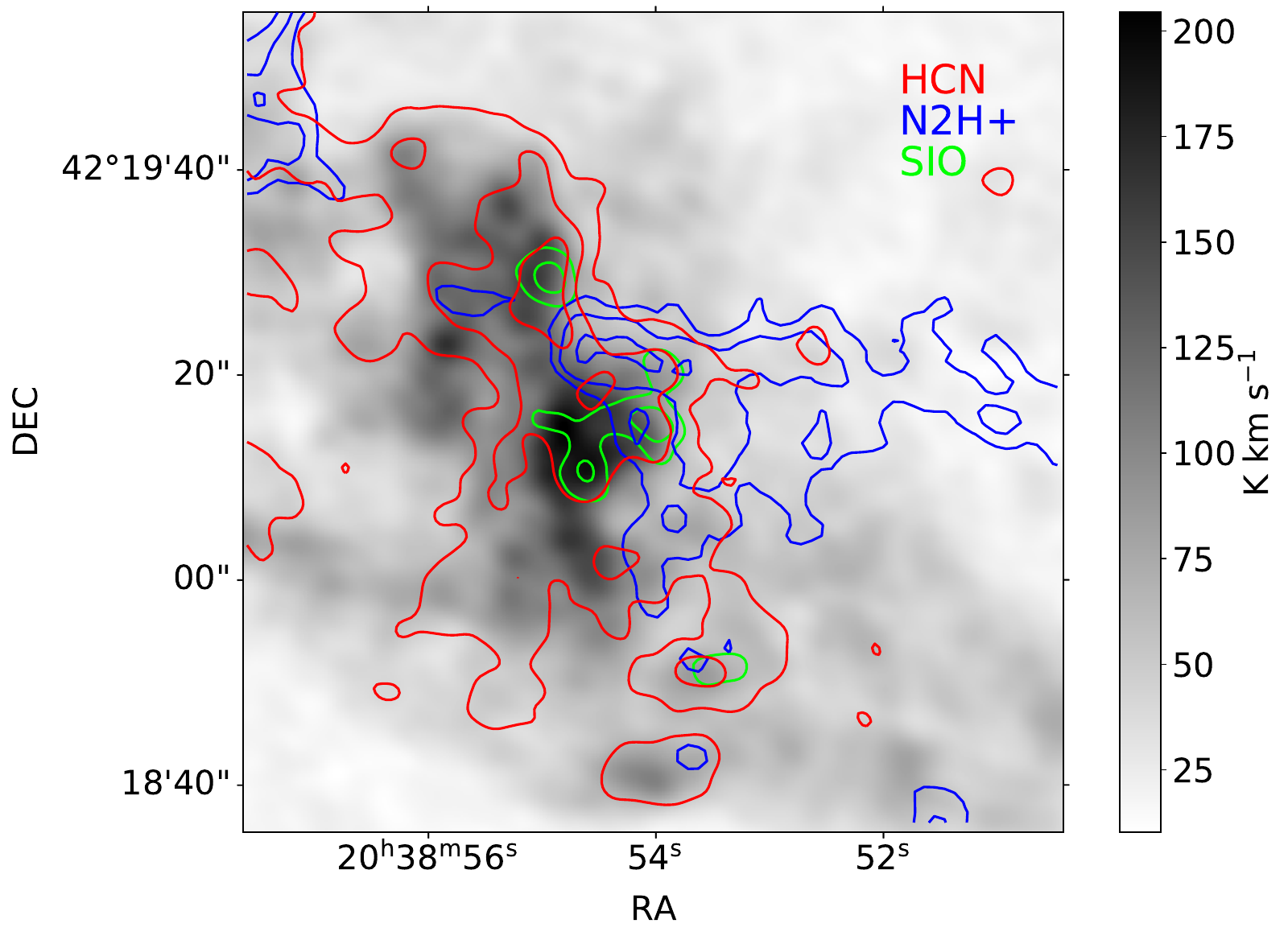}
    \includegraphics[width=8cm]{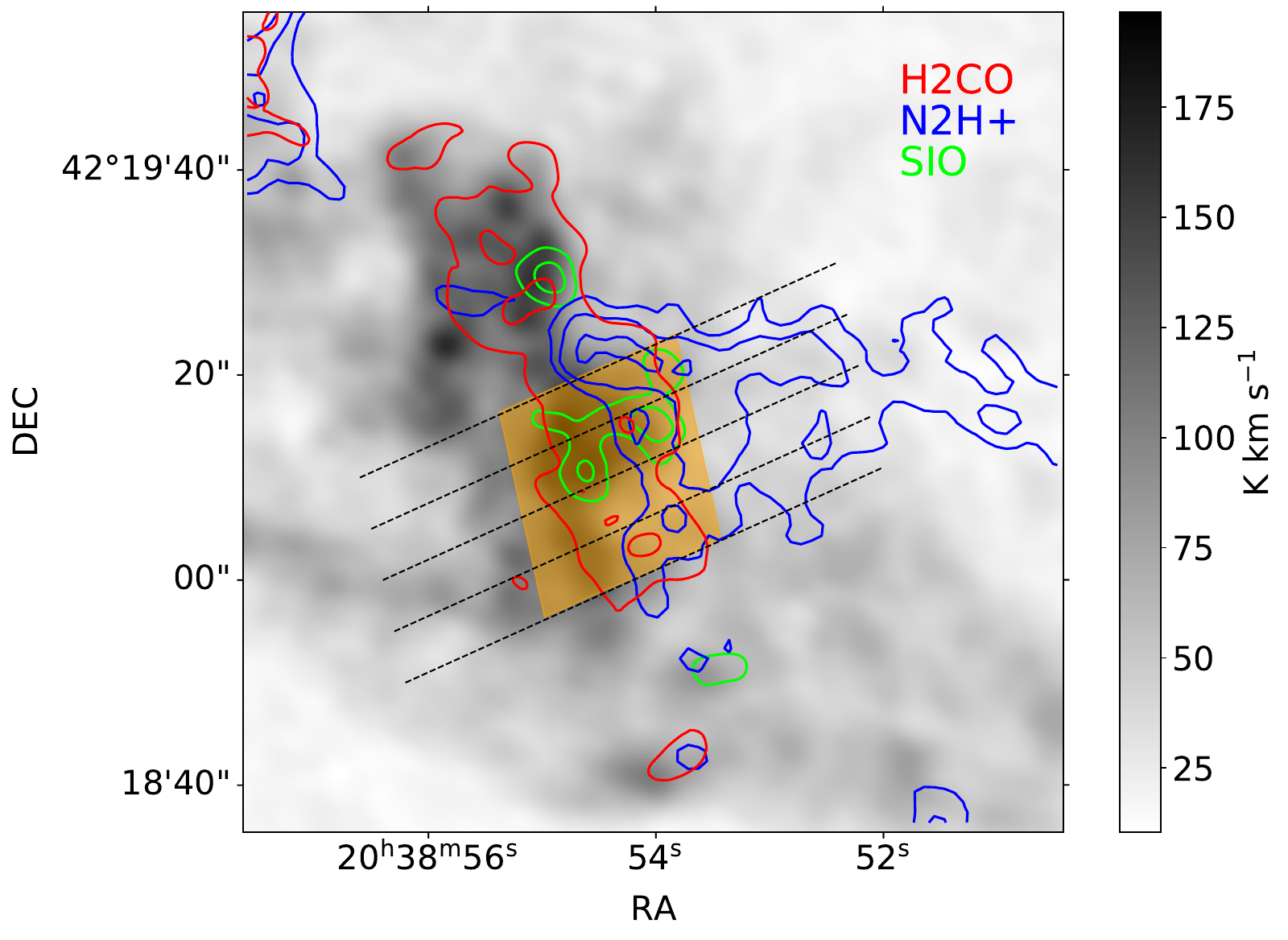}
    \includegraphics[width=8cm]{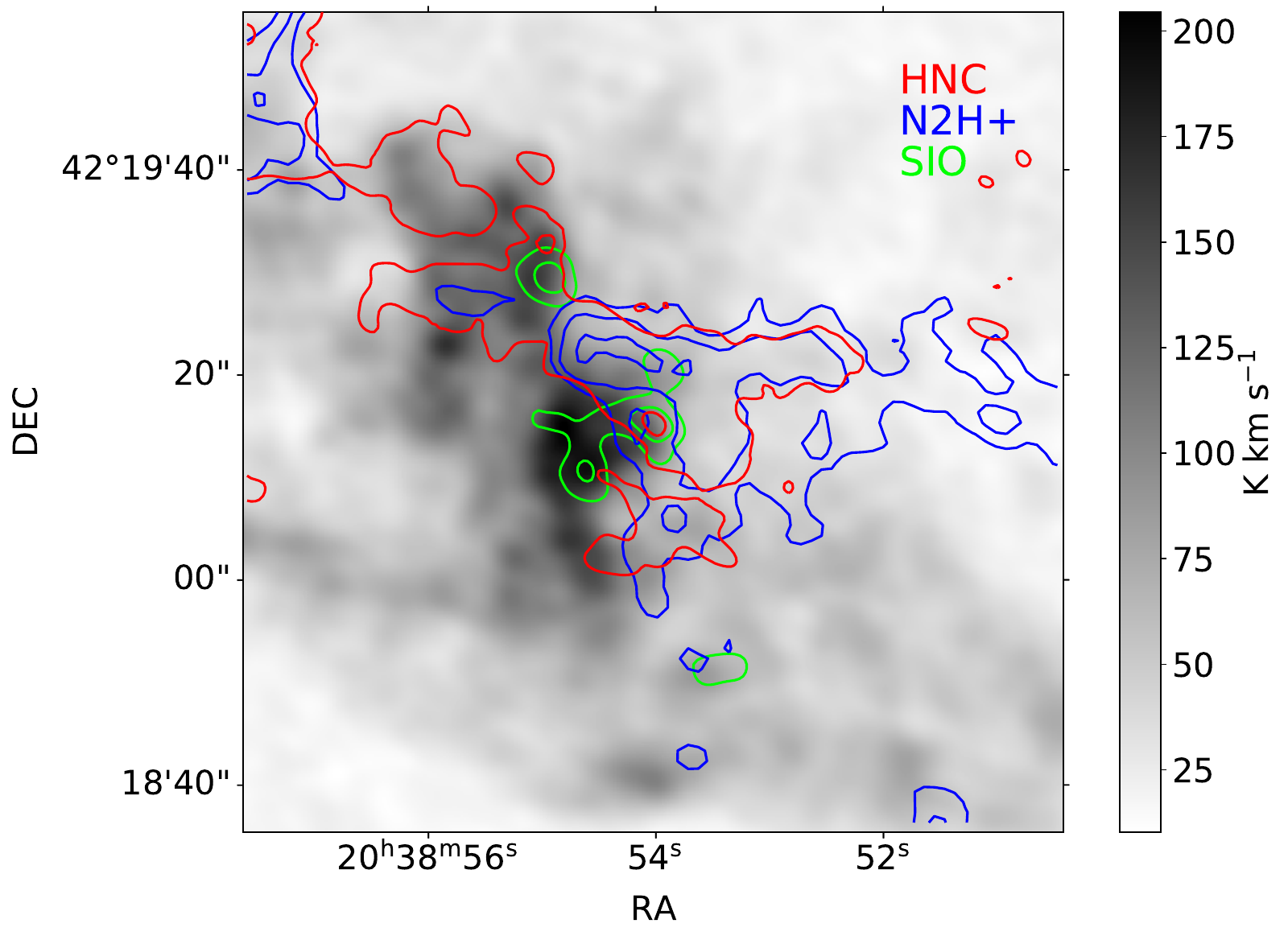}
    \caption{Outflow-cloud interaction region in the western lobe of the DR21 Main outflow. The grey-scale shows the integrated HCO$^+$ emission between $-$50 and 50 km s$^{-1}$ relative to the source velocity. Red contours show line emission of HCN (top), H$_2$CO (middle), and HNC (bottom). Blue contours show the emission of N$_2$H$^+$ and green contours show the emission of SiO, in all panels. Contour levels are at 5, 10 and 20 $\sigma_\text{rms}$. The dashed black lines in the middle plot mark the lines used to calculate the average intensities and first moments across the interaction front, presented in Figs. \ref{fig:intens_across_line}-\ref{fig:HCCCN_CCH}. The orange rectangle marks the area actively affected by the interaction as derived from these intensities.}
    \label{fig:interactioncontours}
\end{figure}
% Intensities across the interaction area figures
\begin{figure}
    \centering
    \includegraphics[width=8cm]{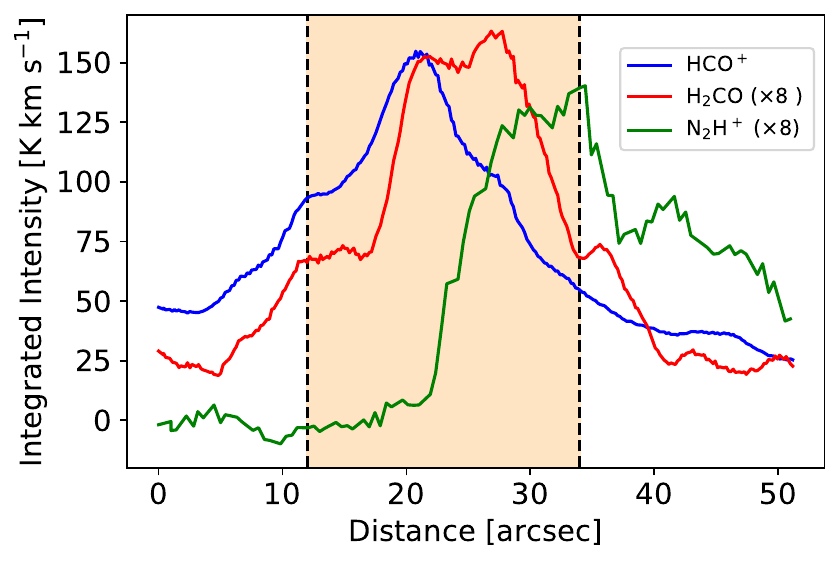}
    \caption{Average integrated intensities of HCO$^+$ (in blue), H$_2$CO (in red), and N$_2$H$^+$ (in green) across the interaction region in the western lobe of DR21. Intensities are integrated from $-$70 to 70, $-$20 to 20 and $-$20 to 10 km s$^{-1}$ for HCO$^+$, H$_2$CO and N$_2$H$^+$, respectively. The x-axis shows the distance in arcseconds, covering the extent of the relevant region where the outflow interacts with a dense structure (marked also in Fig. \ref{fig:interactioncontours}). The orange rectangle shows the area actively affected by the interaction. The intensities for H$_2$CO and N$_2$H$^+$ are scaled up by a factor of 8 in order for their distributions to be more easily comparable.}
    \label{fig:intens_across_line}
\end{figure}

\begin{figure}
    \centering
    \includegraphics[width=8cm]{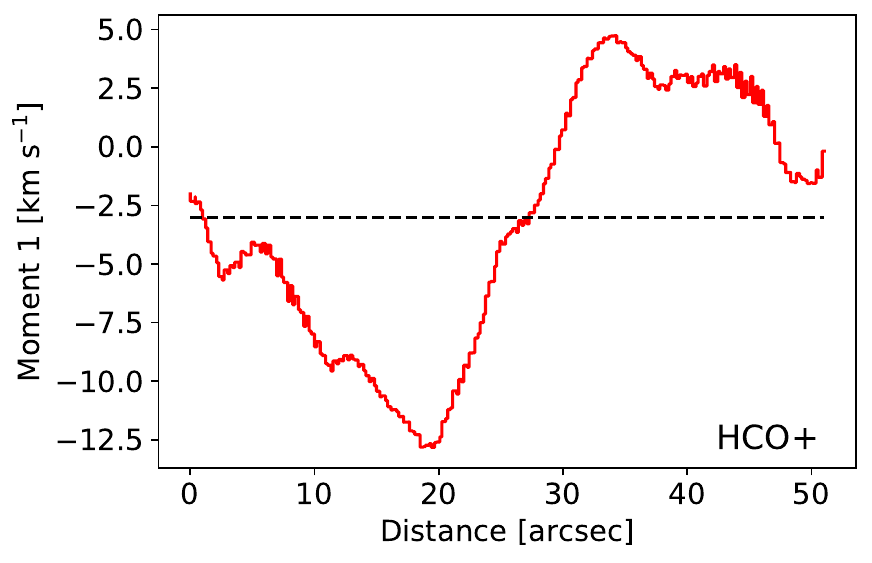}
    \caption{First moment of HCO$^+$ emission across the interaction front (see Fig. \ref{fig:intens_across_line}). The black dashed line marks the DR21 cloud velocity of v$_\text{cloud}$ = $-3$ km s$^{-1}$.}
    \label{fig:HCO+moment1}
\end{figure}

\begin{figure}
    \centering
    \includegraphics[width=8cm]{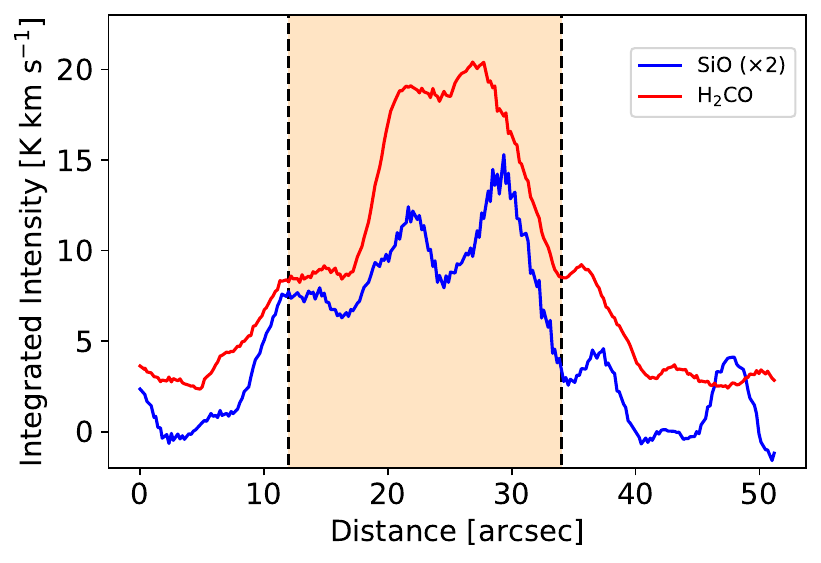}
    \caption{Average integrated intensities of SiO (in blue) and H$_2$CO (in red) across the interaction region in the western lobe of DR21. Intensities are integrated from $-$25 to 25 and $-$20 to 20 km s$^{-1}$ for SiO and H$_2$CO respectively. The SiO  emission is scaled up by a factor of 2 for clarity.}
    \label{fig:H2COvsSiO}
\end{figure}

\begin{figure}
    \centering
    \includegraphics[width=8cm]{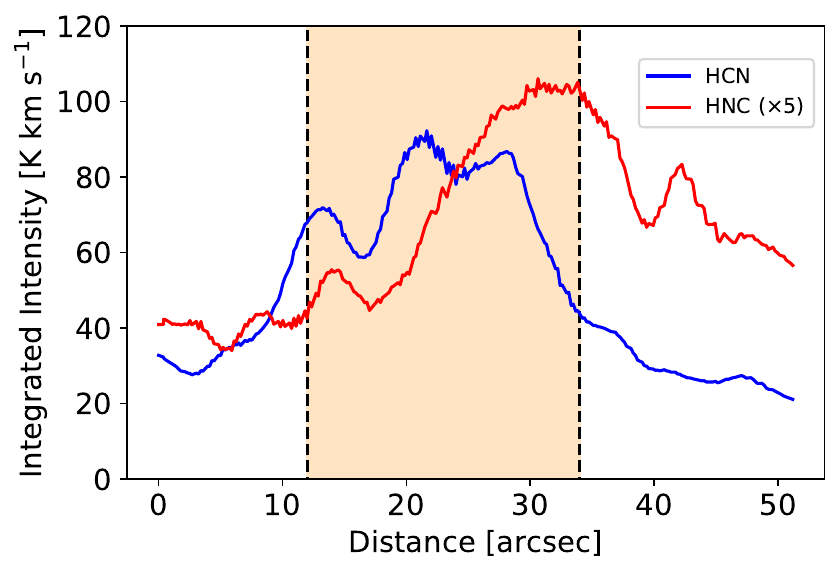}
      \includegraphics[width=8cm]{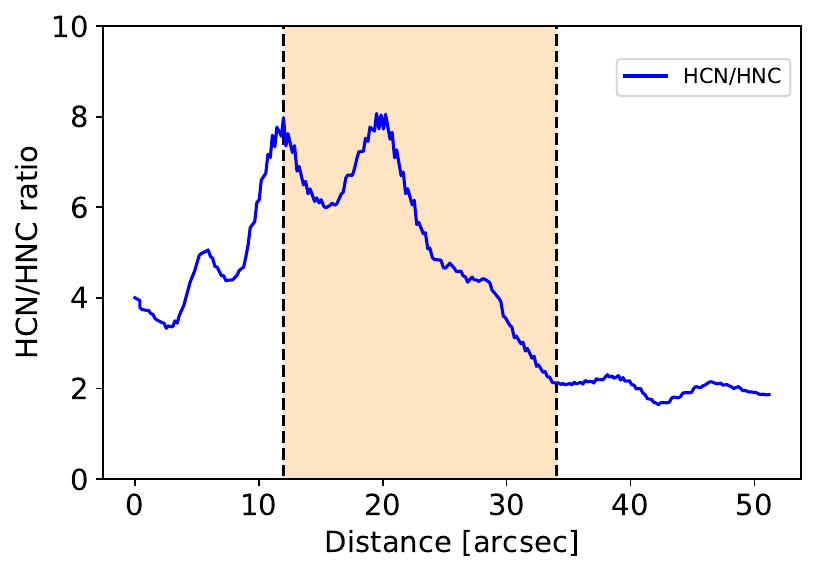}
    \caption{HCN and HNC emission in the interaction region. \textbf{Top}: Average integrated intensities of HCN (in blue) and HNC (in red) across the interaction region in the western lobe of DR21. Intensities are integrated from $-$35 to 35 and $-$15 to 10 km s$^{-1}$ for HCN and HNC respectively. The HNC emission is scaled up by a factor of 5 for clarity. \textbf{Bottom}: Ratio of HCN over HNC across the interaction region.}
    \label{fig:HCNvsHNC}
\end{figure}

Outflow activity from protostars can have an impact on the structure and chemistry of their parental, or nearby, dense cores \citep[e.g.,][]{vankempen2009,lis16,Kahle2022}. The energetic outflow from DR21 Main heavily interacts with its surrounding, and creates the \lq\lq interaction region'' in the western outflow lobe, which has been associated with a collisionally excited Class I methanol maser \citep{Plambeck1990}. Molecular line emission from CASCADE allows us to pin-point the detailed characteristics of the outflow-cloud interaction. 

The interaction region shows various patterns of molecular line emission with HCN and H$_2$CO peaking in its east part, and HNC and N$_2$H$^+$ emission extending to the west and outer part of the outflow lobe (Fig. \ref{fig:interactioncontours}). The SiO emission shows a compact pattern associated most closely with H$_2$CO, suggesting the presence of shocks in the interface between e.g., HCN and N$_2$H$^+$ gas. 

\begin{figure}[h]
    \centering
    \includegraphics[width=8cm]{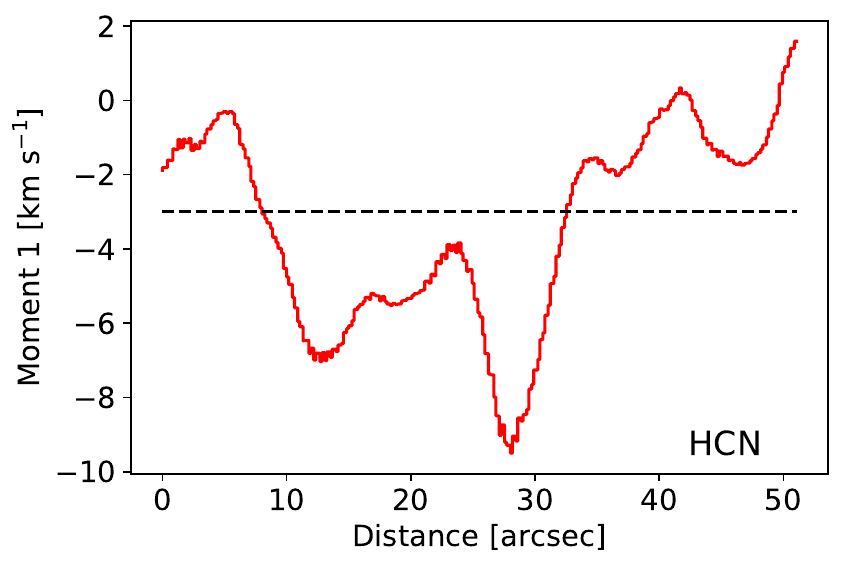}
    \includegraphics[width=8cm]{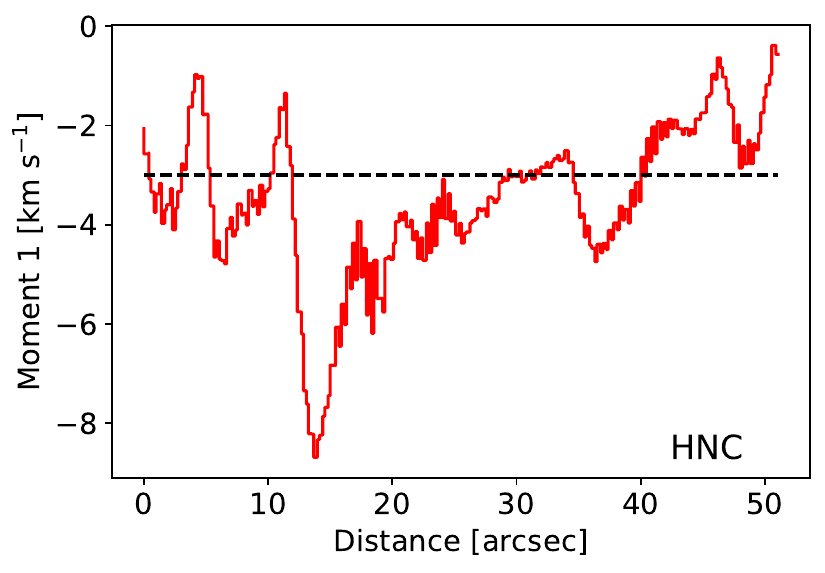}
    \caption{\textbf{Top}: First moment of HCN emission across the interaction front of the DR21 Main outflow and the dense structure located near the western outflow lobe plotted over the corresponding distance. The distance is measured from the outflow dominated part and extends into the dense structure. The black dashed line marks the DR21 cloud velocity of v$_\text{cloud}$ = $-3$ km s$^{-1}$. \textbf{Bottom:} Same as above for but for HNC.}
    \label{fig:HCN_moment1}
\end{figure}

\begin{figure}[h]
    \centering
    \includegraphics[width=8cm]{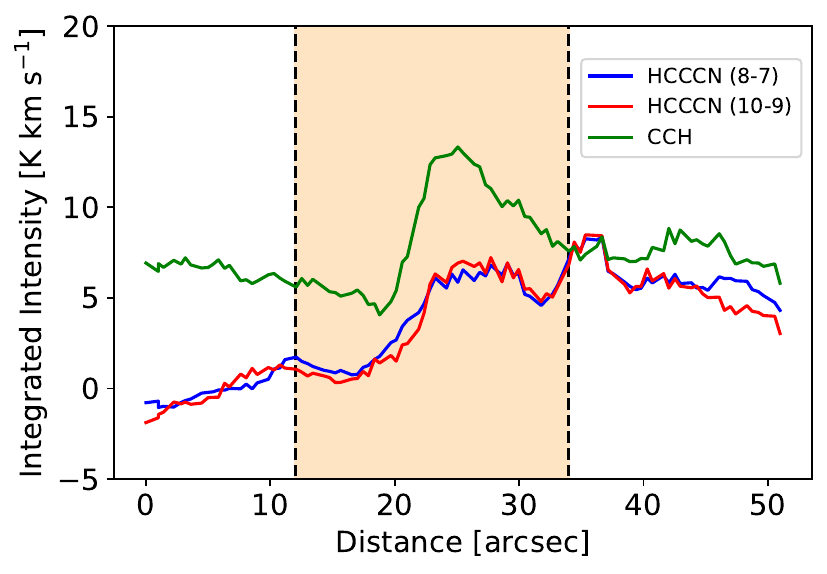}
    \caption{Average integrated intensities of HCCCN $J = 8-7$ (in blue), HCCCN $J = 10 - 9$ (in red), and CCH (in green) across the interaction region in the western lobe of DR21. Intensities are integrated from -10 to 5km s$^{-1}$ for all three lines.}
    \label{fig:HCCCN_CCH}
\end{figure}

We investigate the line emission across the interaction regions in Figures \ref{fig:intens_across_line}-\ref{fig:HCCCN_CCH}. HCO$^+$ emission peaks in the interaction region, and is followed by H$_2$CO tracing warm gas and N$_2$H$^+$, which is sensitive to the sharp increase in cold gas density (Fig.~\ref{fig:intens_across_line}). In the case of HCO$^+$, it is likely that some outflowing material is deflected into the line of sight, giving rise to the strong, high velocity, blue-shifted emission detected in this area (Fig. \ref{fig:velsteps}). This velocity shift is also clearly seen in the first moment of HCO$^+$, which shows that its emission in the area of the dense structure is mainly red-shifted (Fig.~\ref{fig:HCO+moment1}). A likely explanation could be that the dense structure is located closer to the observer along the line of sight and is therefore interacting mostly with the blue-shifted part of the outflow. 
The blue-shifted emission indeed shows a high velocity components elongated almost vertical to the outflow axis, while the red-shifted emission appears almost unperturbed (Fig.~\ref{fig:velsteps}).

The H$_2$CO emission displays a sharp increase in the interaction area, followed by a significant decrease deeper into the dense structure (also shown in Fig. \ref{fig:intens_across_line}). Such an increase is likely related to the enhanced gas temperatures in the interaction region, which lead to the sputtering of H$_2$CO from the dust grains \citep{Benedettini2013}. Alternatively, the H$_2$CO emission could be explained by shock chemistry \citep[e.g.][]{Viti2011}, but the commonly predicted double-peaked structure is not resolved in our observations. It is possible, however, that averaging along the interaction front leads to the blending of line emission making this feature less apparent. A close association of H$_2$CO with the SiO emission tracing shocks (Fig.~\ref{fig:H2COvsSiO}), favors the scenario that H$_2$CO traces not only warm gas, but also the location of active shocks as was suggested by \citet{li2022}. The multiple peaks of SiO and H$_2$CO likely arise due to averaging across the entire interaction front and the clumpy nature of SiO emission (see Fig. \ref{fig:interactioncontours}).

HCN and HNC show some differences along the interaction region (Fig. \ref{fig:HCNvsHNC}), which might reflect the changes in gas temperature \citep{Hacar2020}. The pattern of emission in HNC is similar to that of N$_2$H$^+$, whereas HCN follows closely H$_2$CO. We refrain from using the \citet{Hacar2020} relation to calculate gas temperatures, because the HCN over HNC line ratios partly exceed the range where the experimental relation hold. Nevertheless, the ratio of the two species suggests that the temperature increases rapidly at the front of the interaction area and then drops steadily to a relatively low value. The ratio shows two peaks that follow the peaks of the HCN intensity (Fig. \ref{fig:HCNvsHNC}), and are found close to but not exactly at the same location as the peaks of SiO. This suggests that HCN might be also enhanced in the warm gas behind the shock front \citep[see e.g.,][]{mirocha21}. In addition, velocities of the gas traced by HCN and HNC also differ (Fig. \ref{fig:HCN_moment1}). HNC has velocities close to the cloud velocity, especially from the middle of the interaction area and into the dense structure, where its emission actually becomes significant. HCN on the other hand follows HCO$^+$ and shows significant blue-shifted emission, which becomes red-shifted in the area dominated by the dense structure. The blue-shifted peak appears deeper in the interaction area compared that of HCO$^+$ showing that they trace different material.  

HCCCN emission is detected in the area of the dense structure as expected based on previous studies \citep[e.g.][]{Morris1976,churchwell1978}. The slight enhancement in the interaction region suggest the origin in shocks (Fig. \ref{fig:HCCCN_CCH}), in agreement with \cite{Benedettini2013}.

The intensity of CCH shows a significant increase close to the center of the interaction region, likely due to UV radiation originating from the shocked material in the interaction region \citep[e.g.][]{Gratier2017,Bouvier2020,Chahine2022}.

In contrast, HNCO is not detected in the interaction region, even though it is often associated with shocks. According to \citet{Yu2018}, HNCO is preferentially enhanced in slow moving shocks, but destroyed in high velocity shocks, which might indeed be the case for the energetic outflow from DR21.

Finally, we note here that due to the irregular shape of the interaction front, the simplistic average along it adopted for the above analysis suffers from significant uncertainties. Still, it can offer an interesting qualitative look into the behaviour of species across such an interaction. 

%% More interaction plots

% -------- 

\section{Summary and conclusions}
\label{sec:summary}

This work presents the results of the CASCADE observations in the area of the DR21 Main outflow covering several molecular tracers, including  HCO$^+$, HCN, HNC, N$_2$H$^+$, H$_2$CO, and CCH at high spatial ($\sim3\arcsec$) and spectral resolution ($\sim$0.8 km s$^{-1}$). 
These molecular tracers are split into three separate categories according to their morphology, tracing the outflow (HCO$^+$ and HCN), the DR21 ridge ($^{13}$CS, CCH, H$_2$CO, H$^{13}$CO$^+$, HCCCN (10 -- 9), HCCCN (8 -- 7), HNC, N$_2$H$^+$) and localized emission, e.g. CH$_3$CN (4$_k$ -- 3$_k$), CH$_3$CN (5$_k$ -- 4$_k$), CH$_3$OH, DCN, DCO$^+$, DNC, H$\alpha$41, NH$_2$D, SiO.

Based on the HCO$^+$ emission, the DR21 Main outflow was found to mostly resemble a typical bipolar outflow rather than an explosive one, as its emission shows two well structured lobes that get progressively more collimated at higher velocities and lack the filament-like structures that are prevalent in established explosive outflows. 

Adapting the separation method, and applying it to HCO$^+$ emission allowed for the estimation of the energetic parameters (outflow force $F$ = 0.56 M$_\odot$ km yr$^{-1}$ s$^{-1}$, mass $M$ = 124 M$_\odot$ and kinetic energy $E_\text{kin}$ = 5 $\times 10^{47}$ erg) of the outflow. Comparison with other protostellar sources showed that the outflow force of the DR21 Main outflow is about an order of magnitude higher than sources of similar envelope mass. It remains uncertain though whether the outflow of DR21 Main represents an upper limit of typical protostellar outflows or is powered by a different mechanism. 

Finally, a dense molecular structure was detected near the western lobe of the outflow. The detection of SiO and H$_2$CO emission in this area showed that there is ongoing interaction between the outflow and this dense structure. This, in turn, allowed for an analysis of the behaviour of different molecular species across such an interaction and found them to be in good agreement with the results of recent modelling predictions and observations of shocked regions. 

Overall, the results presented in this paper firmly establish the outflow of DR21 Main as one of the most interesting cases of bipolar, protostellar outflows due to its exceptional size and power. Additionally, the CASCADE observations offer a good glimpse into the chemistry of the interaction region but further and more detailed modeling is required in order to properly constrain the chemistry that takes place in this location.

\begin{acknowledgements}
The authors are grateful to the staff at the NOEMA and Pico
Veleta observatories for their support of these observations. We thank in particular P. Chaudet, operator at the NOEMA observatory, for his motivation and dedication in developing and testing the advanced mosaic observing procedures employed in this project. This work is based on observations carried out under project number L19MA with the IRAM NOEMA Interferometer and [145-19] with the 30 m telescope. IRAM is supported by INSU/CNRS (France), MPG (Germany) and IGN (Spain). 
AK acknowledges support from the Polish National Agency for Academic Exchange grant No. BPN/BEK/2021/1/00319/DEC/1. 
H.B. acknowledges support from the European Research Council under the Horizon 2020 Framework Program via the ERC Consolidator Grant CSF-648505 and from the Deutsche Forschungsgemeinschaft in the Collaborative Research Center (SFB 881) “The Milky Way System” (subproject B1).
A.G. acknowledges support from NSF AAG 2008101 and NSF
CAREER 2142300. 
D.~S. acknowledges support from the European Research Council under the Horizon 2020 Framework Program via the ERC Advanced Grant No. 832428-Origins. 
\end{acknowledgements}

\bibliographystyle{aa}

\bibliography{biblio.bib}

\begin{appendix}
\section{Integrated intensity contour maps}
\label{app:contours}
\begin{figure*}
    \includegraphics[width=18cm]{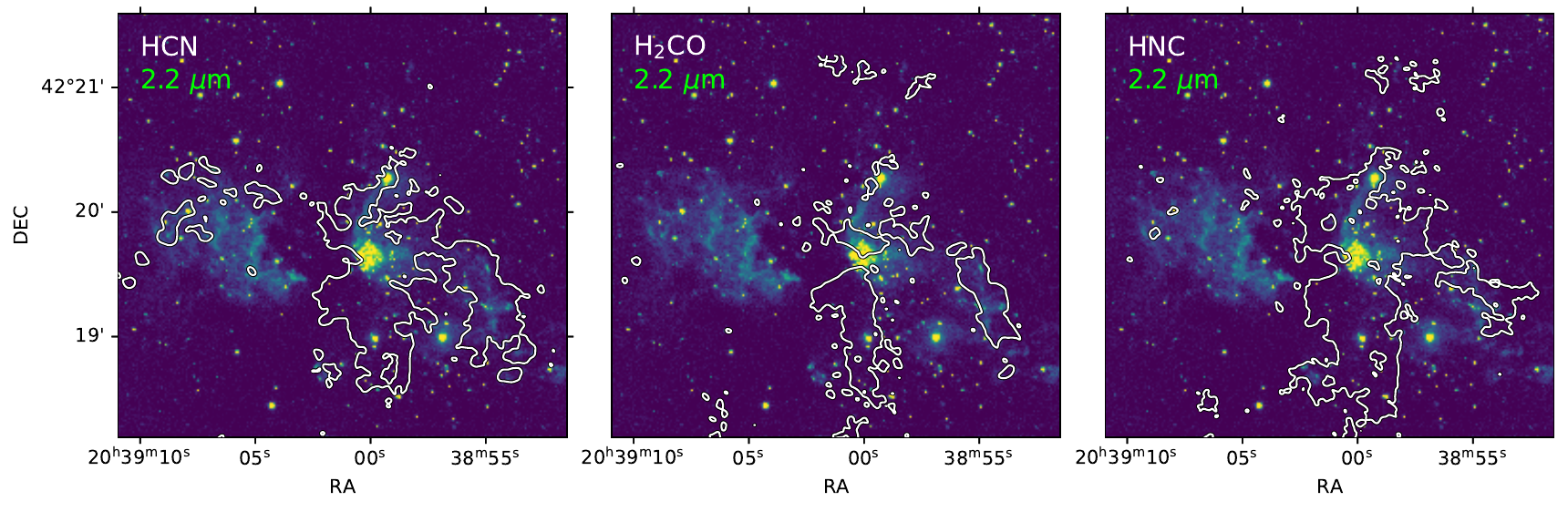}
    \caption{\textit{UKIRT}/WFCAM continuum image of the DR21 Main region at 2.2 $\mu$m and the line emission in key gas tracers observed as part of CASCADE. White contours mark the 5$\sigma$ HCN (left), H$_2$CO (middle) and HNC (right) emission.}
    \label{fig:appcontours1}
\end{figure*}

\begin{figure*}
    \includegraphics[width=18cm]{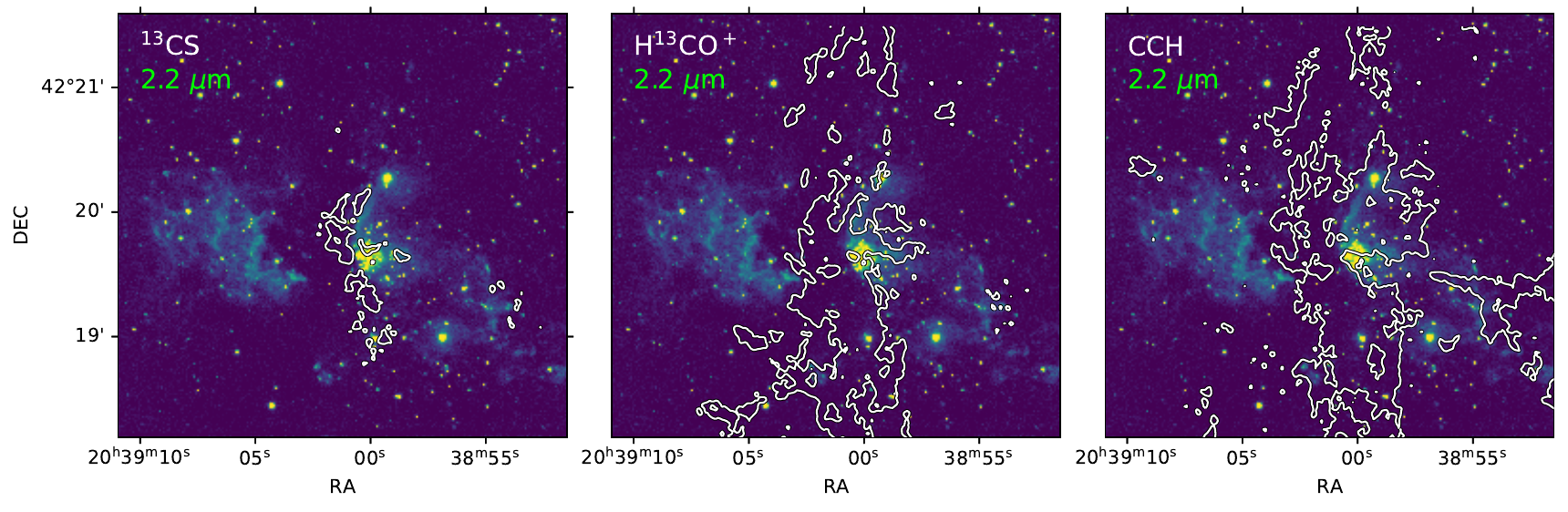}
    \caption{Similar to Fig.~\ref{fig:appcontours1}, but for $^{13}$CS (left), H$^{13}$CO$^+$ (middle) and CCH (right).}
    \label{fig:appcontours2}
\end{figure*}

\begin{figure*}
    \includegraphics[width=18cm]{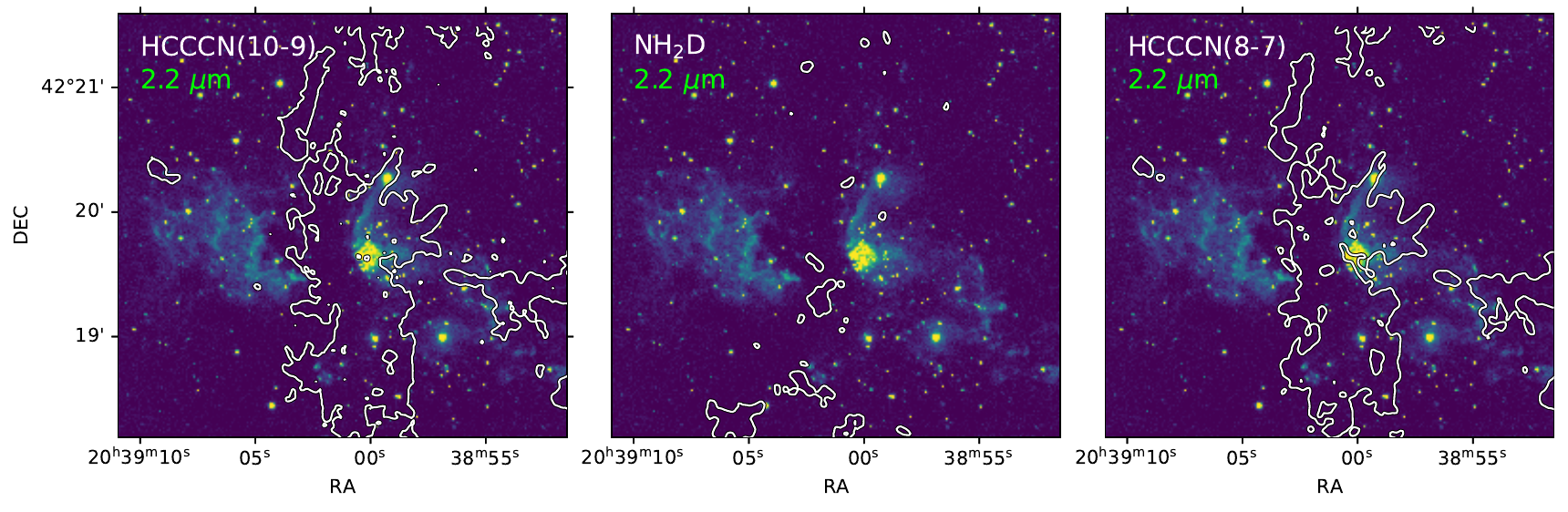}
    \caption{Similar to Fig.~\ref{fig:appcontours1}, but for HCCCN (10-9) (left), NH$_2$D (middle) and HCCCN (8-7) (right).}  
    \label{fig:appcontours3}
\end{figure*}

\begin{figure*}
\includegraphics[width=18cm]{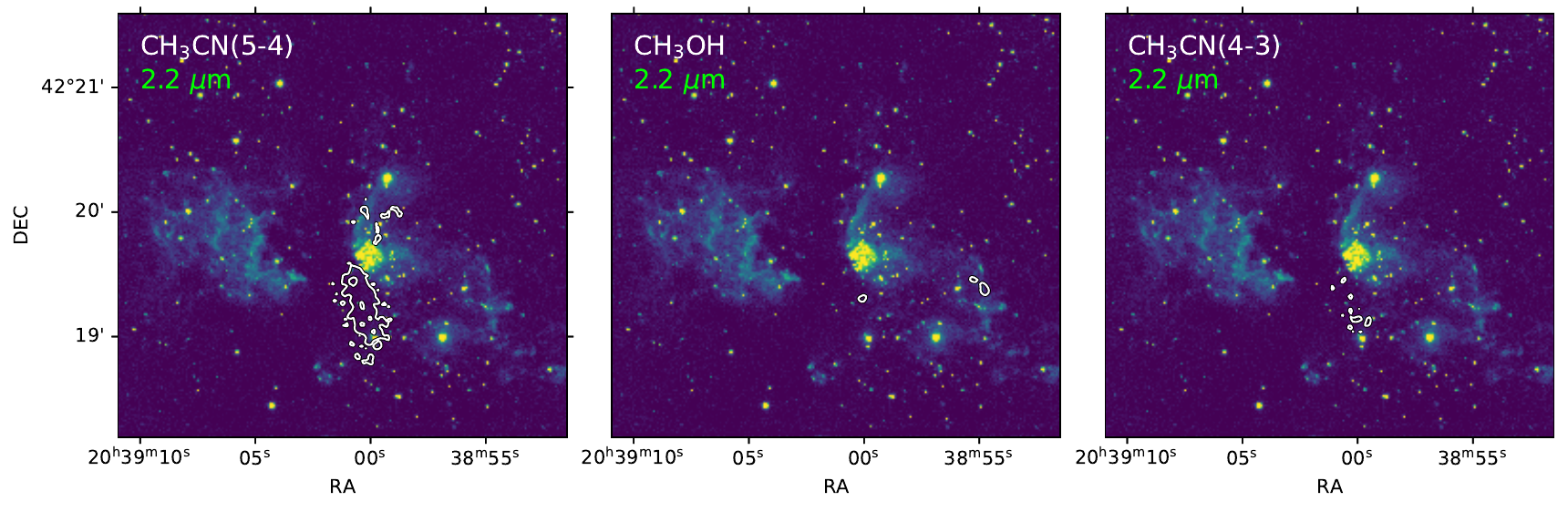}
    \caption{Similar to Fig.~\ref{fig:appcontours1}, for CH$_3$CN (5-4) (left), CH$_3$OH (middle) and CH$_3$CN (4-3) (right).}
    \label{fig:appcontours4}
\end{figure*}

\begin{figure*}
\includegraphics[width=18cm]{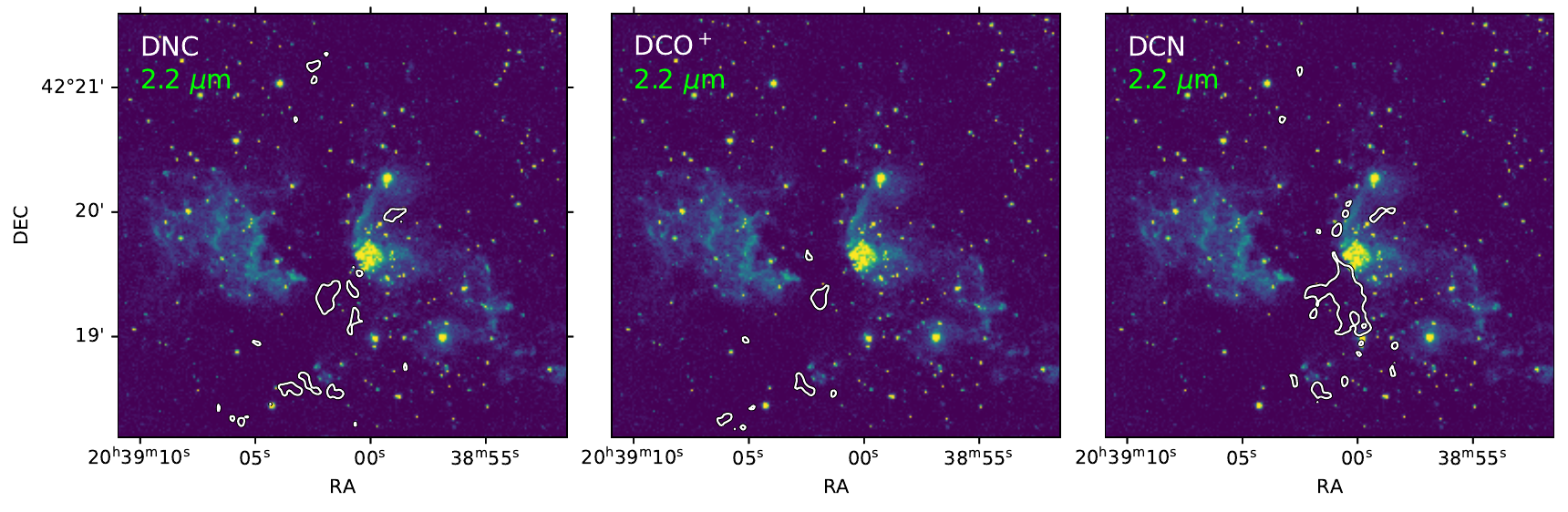}
    \caption{Similar to Fig.~\ref{fig:appcontours1}, but for DNC (left), DCO$^+$ (middle) and DCN (8-7) (right).}
    \label{fig:appcontours5}
\end{figure*}

\begin{figure*}
\includegraphics[width=18cm]{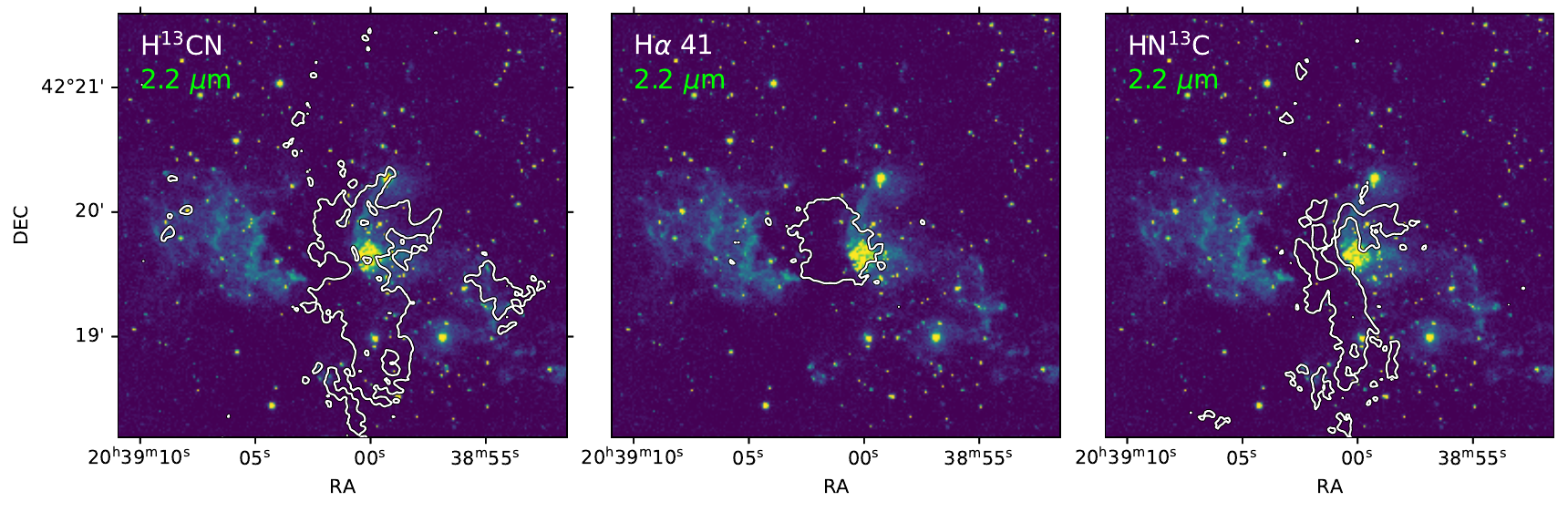}
    \caption{Similar to Fig.~\ref{fig:appcontours1}, but for H$^{13}$CN (left), H$\alpha$41 (middle) and HN$^{13}$C (right).}    
    \label{fig:appcontours6}
\end{figure*}

Figures \ref{fig:appcontours1}--\ref{fig:appcontours6} show contour maps of the integrated intensity of all observed molecules in the area of DR21 Main outflow. The detected molecules can broadly be separated in three categories (see Section \ref{sec:detections}, Table \ref{table:detections}) as i) tracing the outflow, ii) tracing the dense ridge, and iii) displaying localized and fragmented emission. The emission of the rare isotopologues (e.g,  H$^{13}$CO$^+$, H$^{13}$CN, HN$^{13}$C and $^{13}$CS) appears to trace the high density material along the DR21 ridge. In contrast to the patters of emission in H$^{13}$CO$^+$ and H$^{13}$CN, the emission in their $^{12}$C counter-parts, is associated with outflowing material, likely due to low abundances of rare isotopologues in the high-velocity gas.


\FloatBarrier
\section{Line profiles in detected molecules}
\label{app:spectra}
\begin{figure*}
    \centering
    \includegraphics[width=12.5cm]{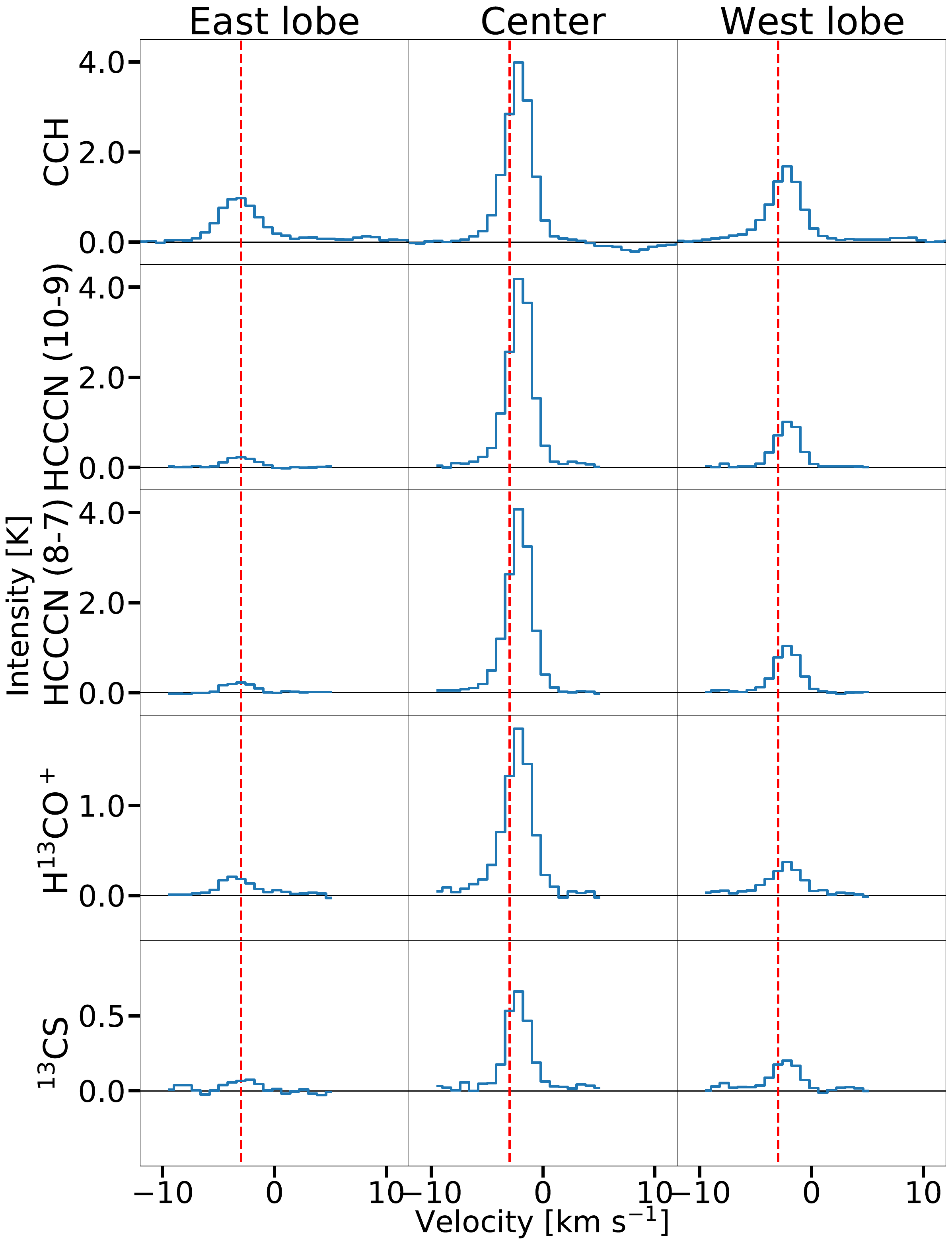}
    \caption{Similar to \ref{fig:allspectra}, but for CCH, HCCCN (10 - 9), HCCCN (8 - 7), H$^{13}$CO and $^{13}$CS }
    \label{fig:allspectra2}
\end{figure*}

\begin{figure*}
    \centering
    \includegraphics[width=12.5cm]{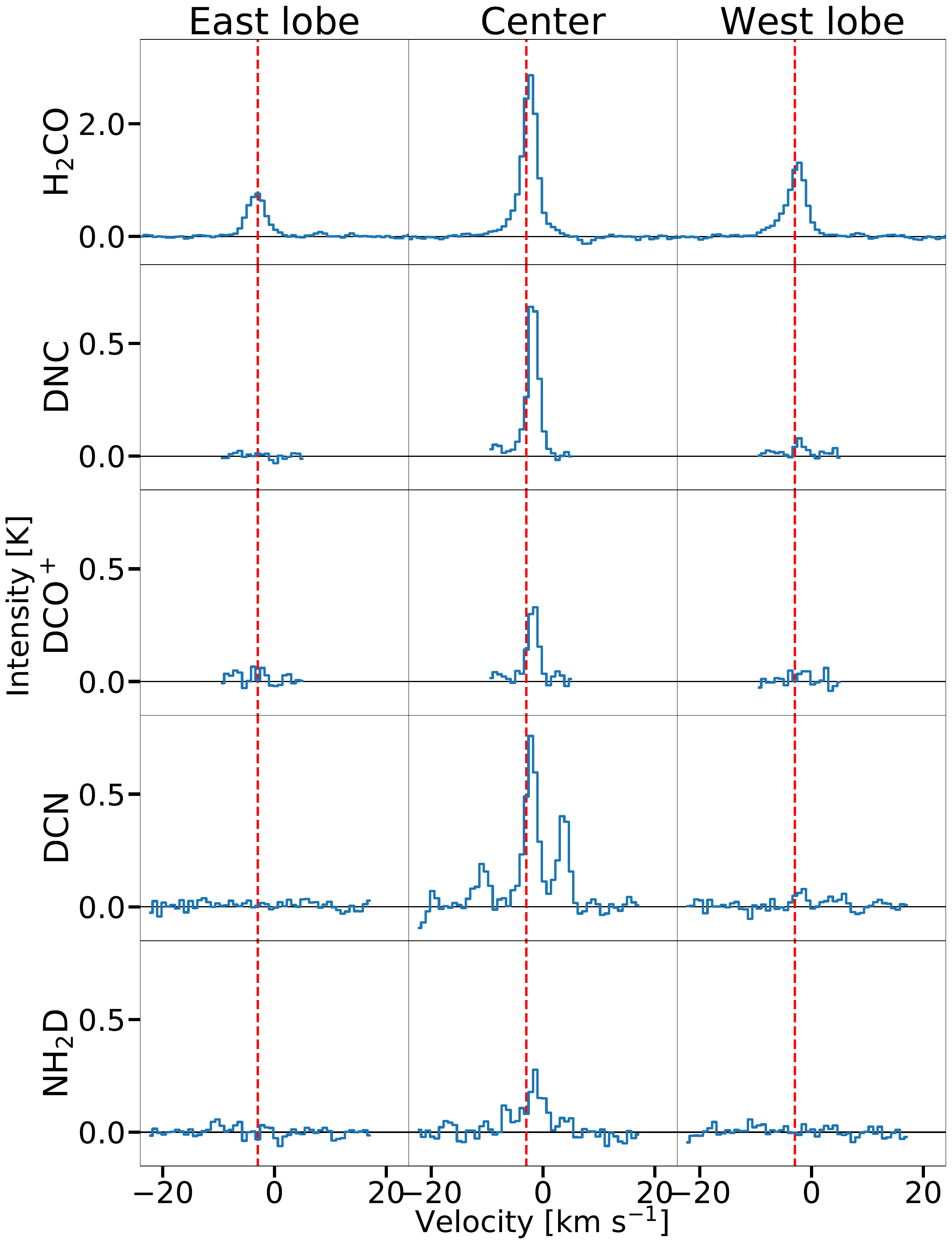}
    \caption{Similar to \ref{fig:allspectra}, but for H2CO, DNC, DCO$^+$, DCN and NH$_2$D}
    \label{fig:allspectra3}
\end{figure*}

\begin{figure*}
    \sidecaption
    \includegraphics[width=12.5cm]{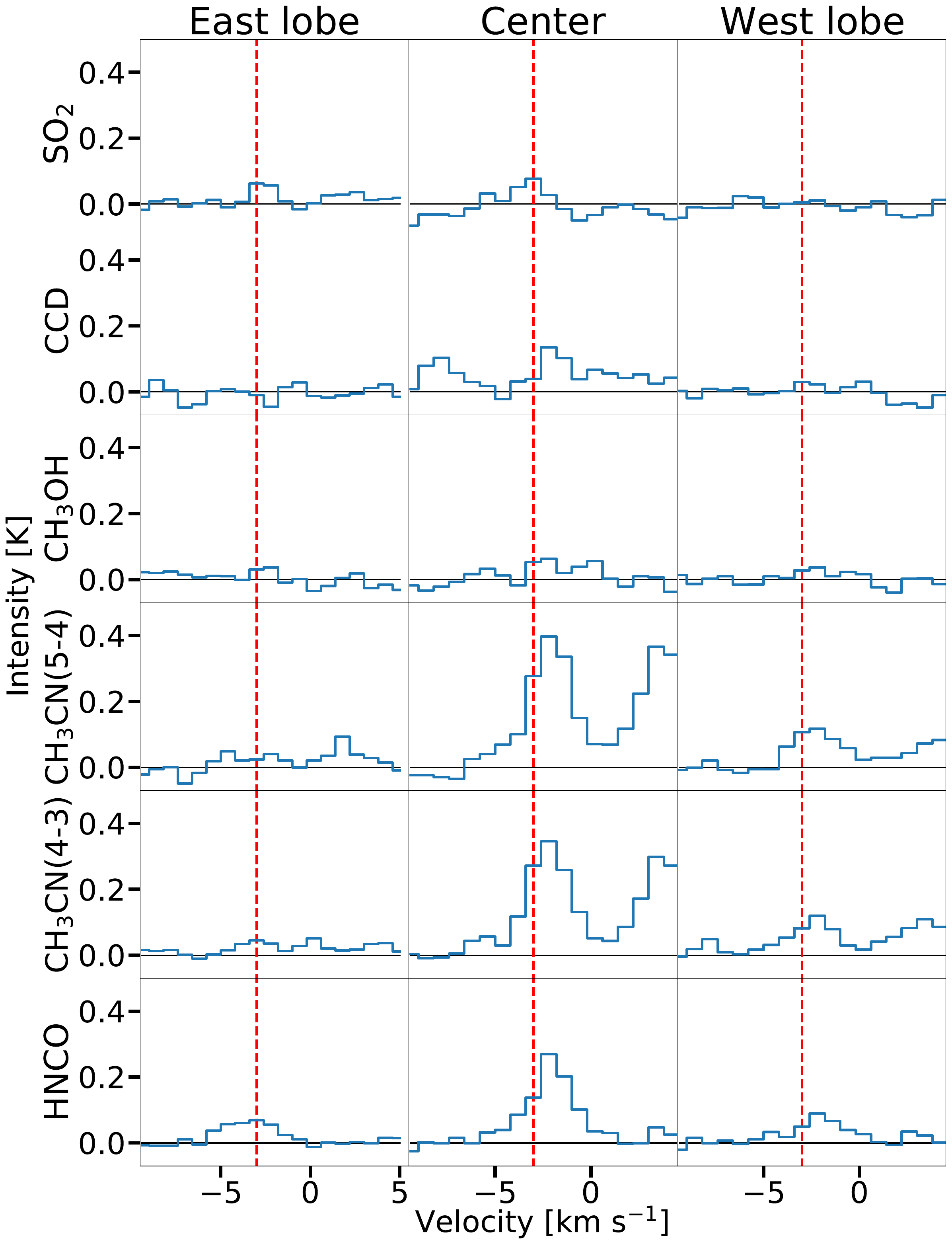}
    \caption{Same as Fig \ref{fig:allspectra}, but for SO$_2$, CCD, CH$_3$OH, CH$_3$CN (4$_k$ - 3$_k$), CH$_3$CN (5$_k$ - 4$_k$) and HNCO }
    \label{fig:allspectra4}
\end{figure*}

\begin{figure*}
    \sidecaption
    \includegraphics[width=12.5cm]{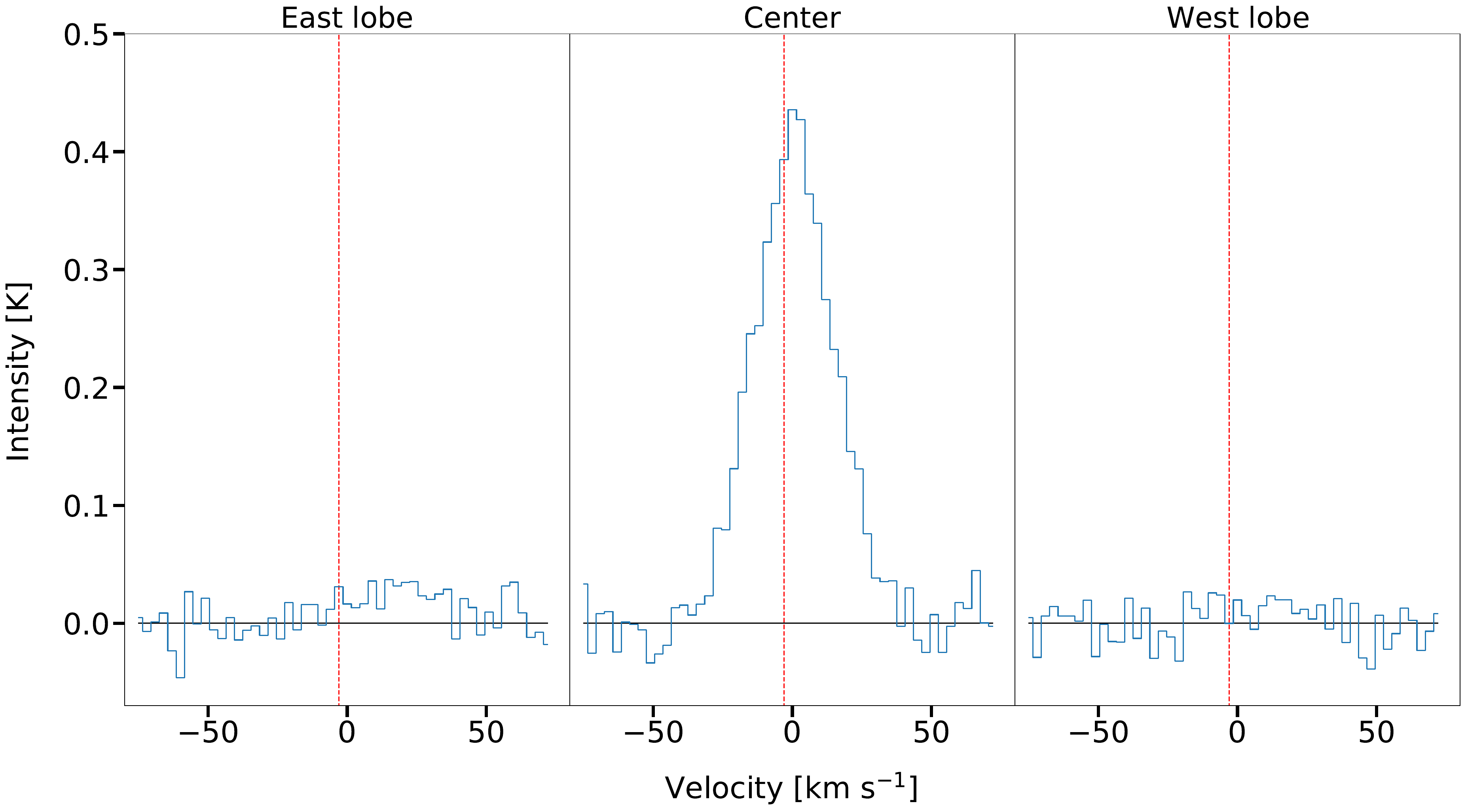}
    \caption{Same as Fig \ref{fig:allspectra}, but for H41$\alpha$ emission.}
    \label{fig:allspectra5}
\end{figure*}

Figures \ref{fig:allspectra2}--\ref{fig:allspectra5} show spectra of the different molecular lines observed in CASCADE. The spectra are averaged over the east lobe, the center and the western lobe (see Fig.~\ref{fig:h2images}).

\FloatBarrier
\section{Correlation plots}
\begin{figure*}
    \centering
    \includegraphics[width=8cm]{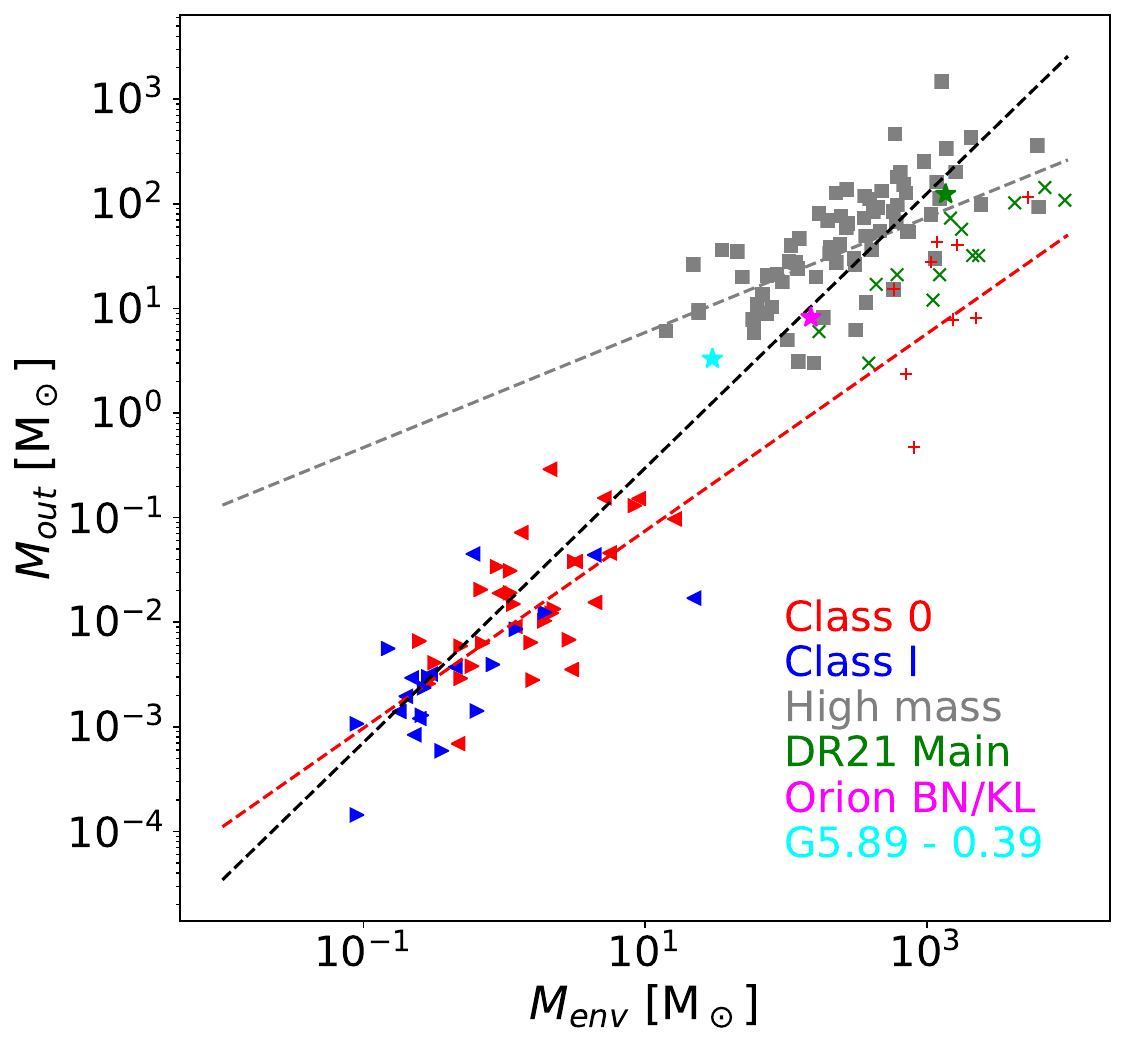}
    \includegraphics[width=8cm]{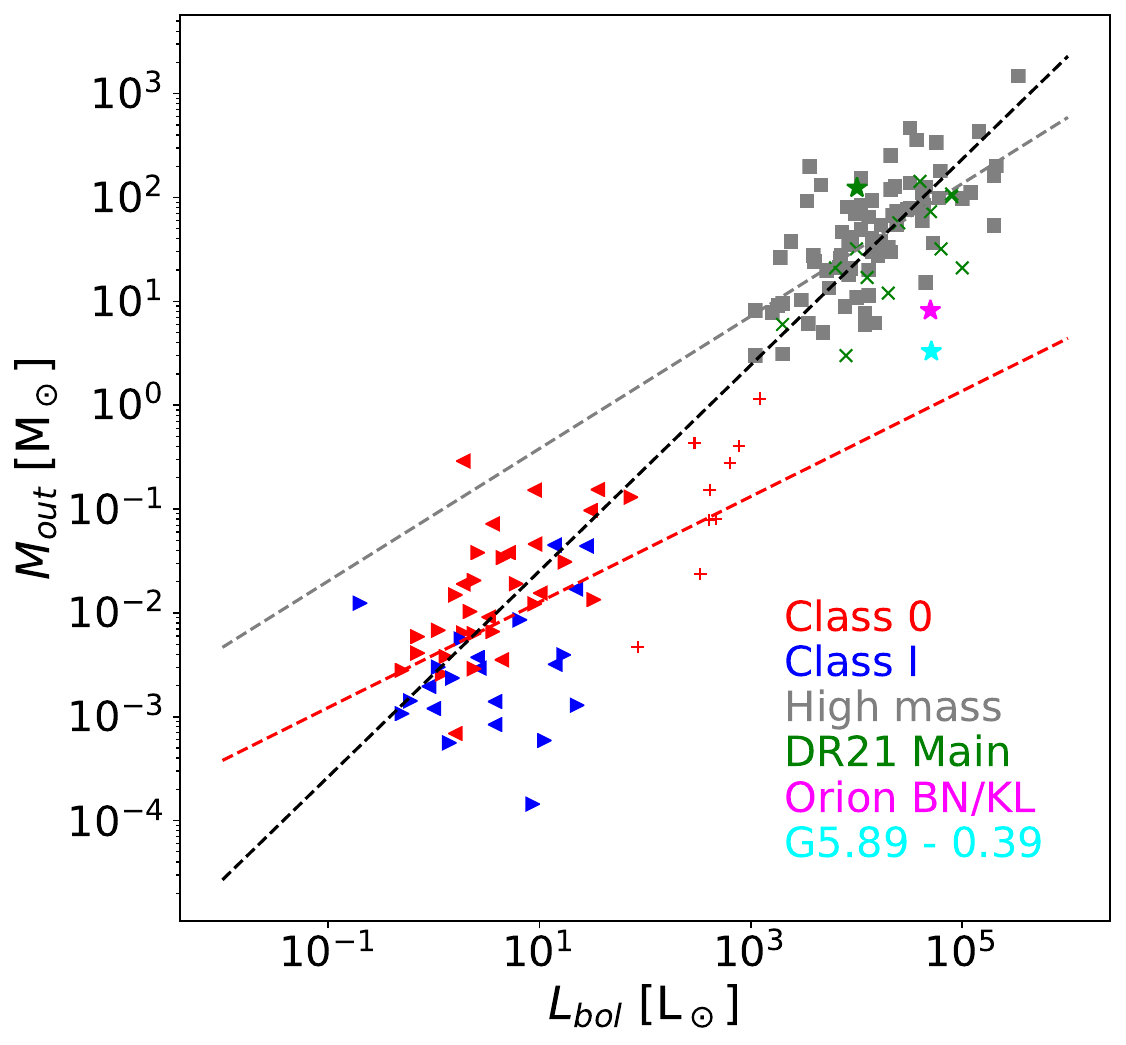}
    \caption{Outflow mass versus envelope mass and bolometric luminosity. The labels are the same as in Fig.~\ref{fig:forcemass}.}
    \label{fig:cor1}
\end{figure*}

\begin{figure*}
    \centering
    \includegraphics[width=8cm]{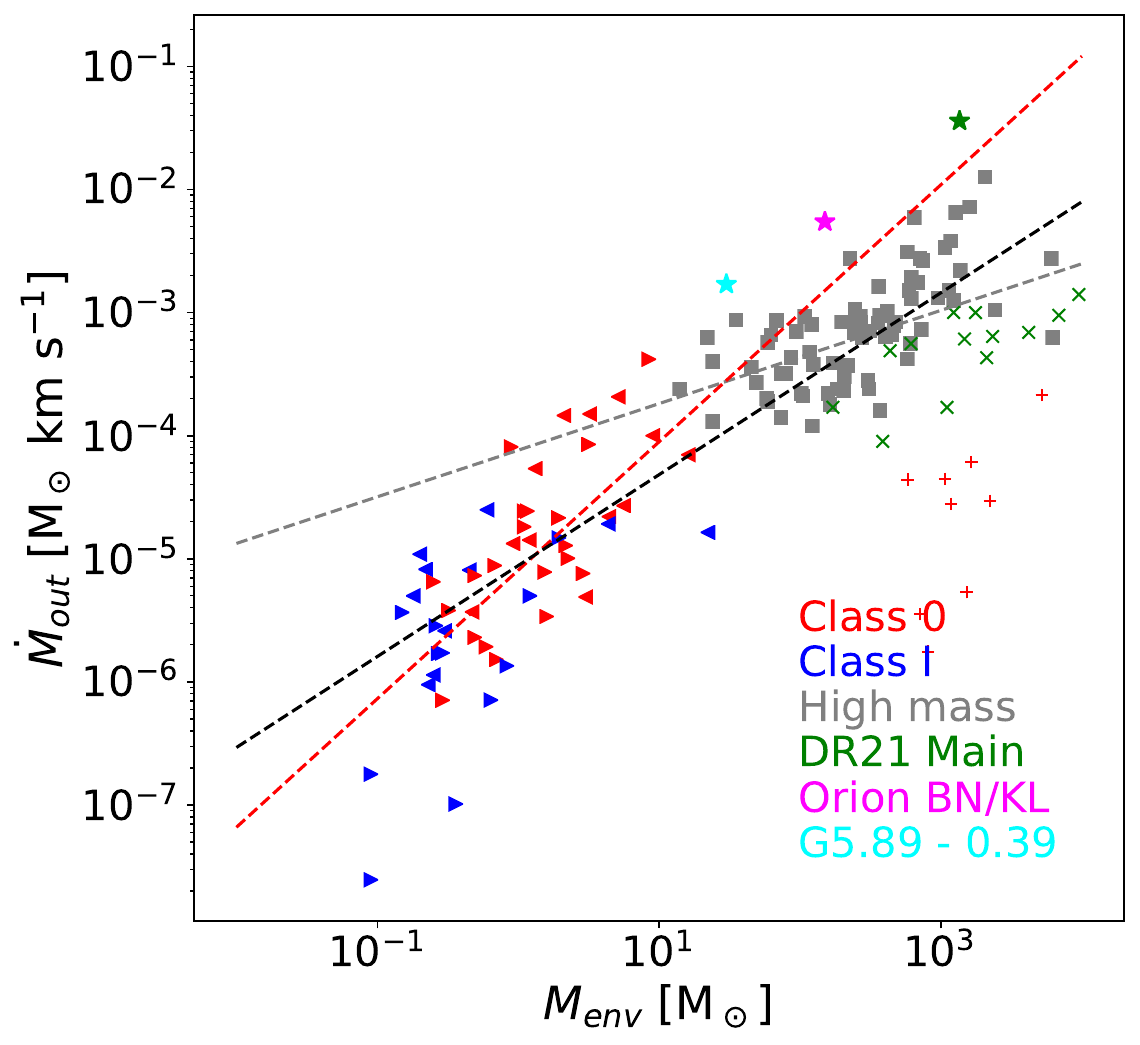}
    \includegraphics[width=8cm]{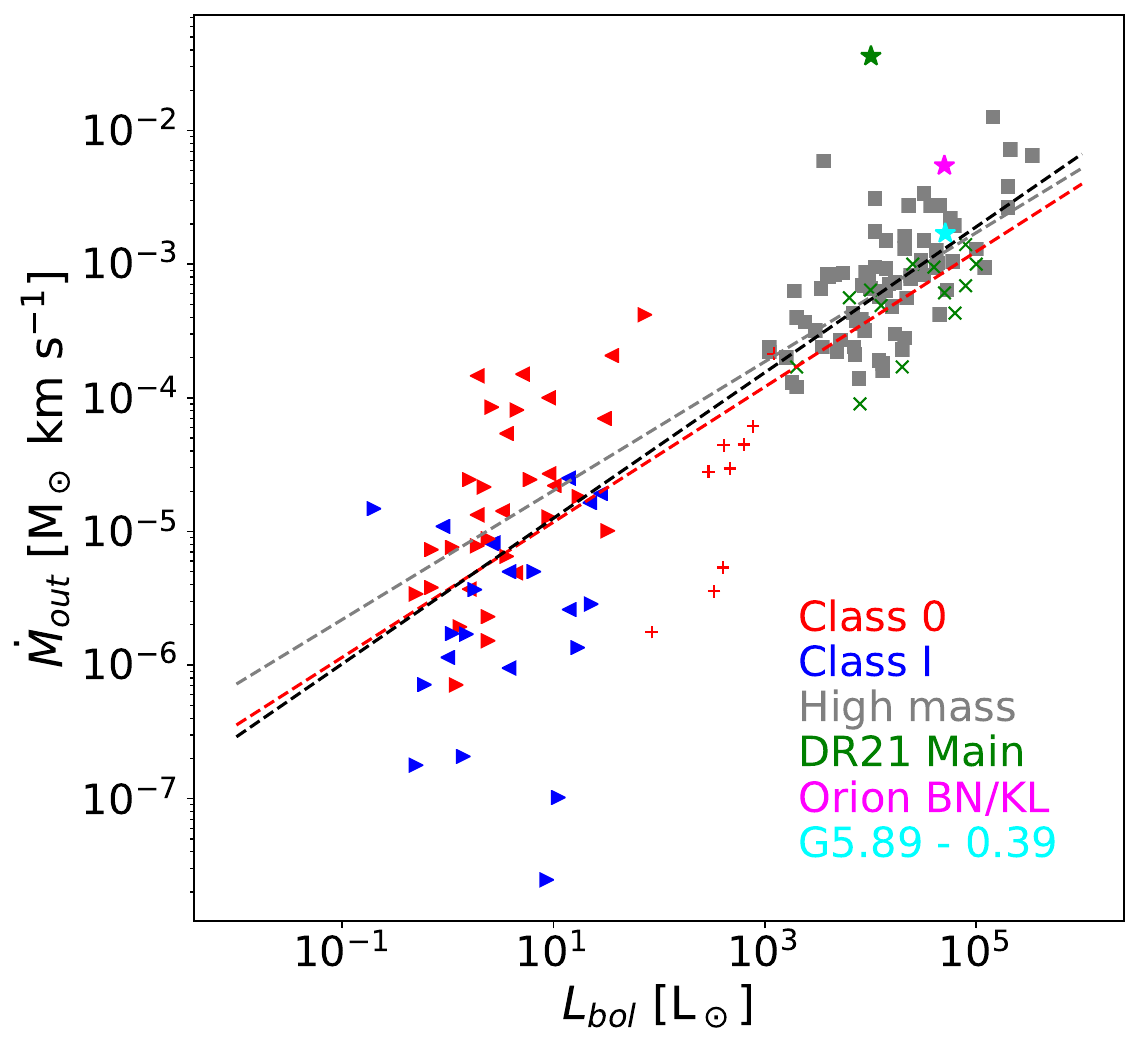}
    \caption{Outflow mass rate versus envelope mass and bolometric luminosity. The labels are the same as in Fig.~\ref{fig:forcemass}.}
    \label{fig:cor2}
\end{figure*}

\begin{figure*}
    \centering
    \includegraphics[width=8cm]{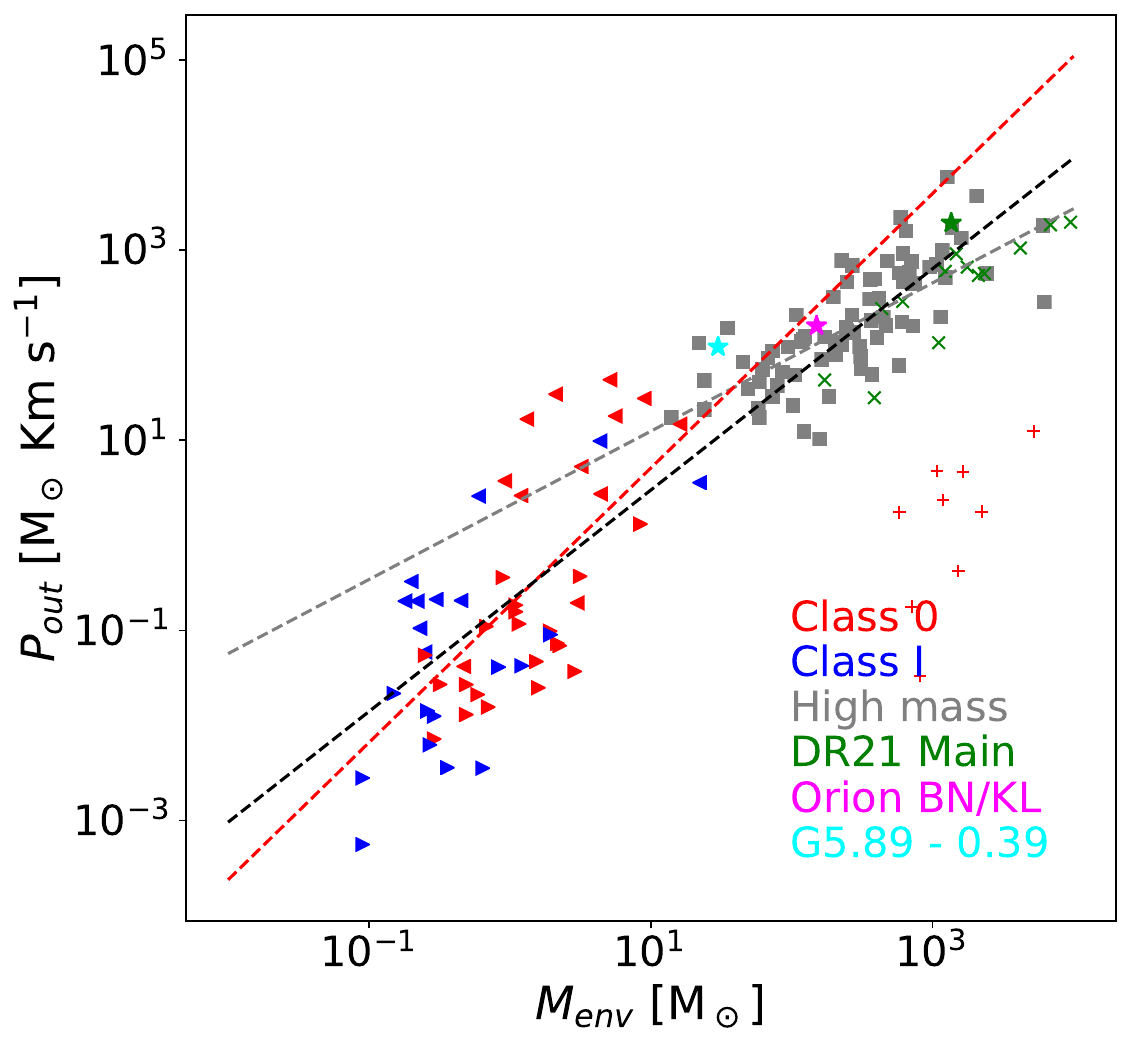}
    \includegraphics[width=8cm]{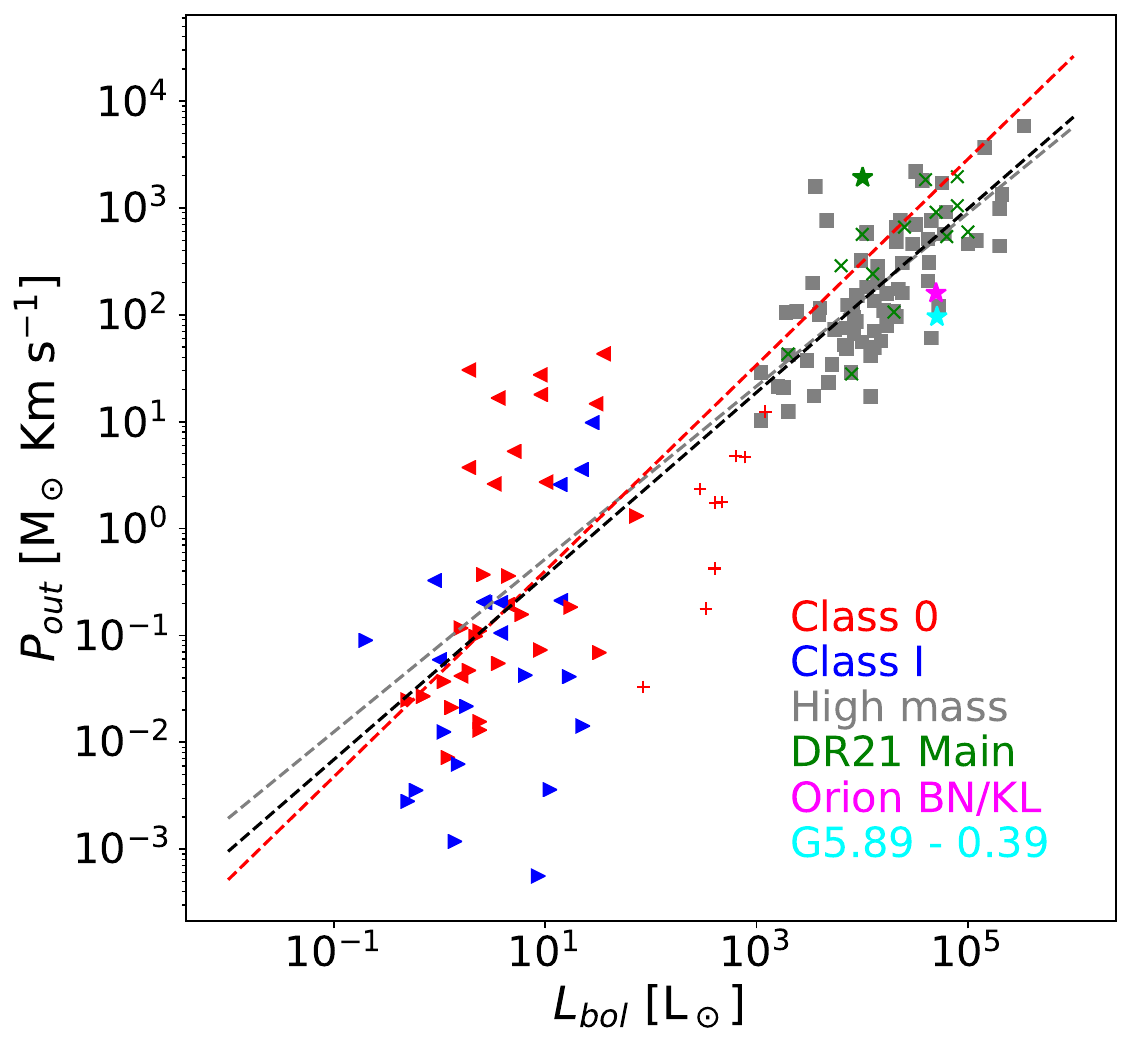}
    \caption{Outflow momentum versus envelope mass and bolometric luminosity. The labels are the same as in Fig.~\ref{fig:forcemass}.}
    \label{fig:cor3}
\end{figure*}

\begin{figure*}
    \centering
    \includegraphics[width=8cm]{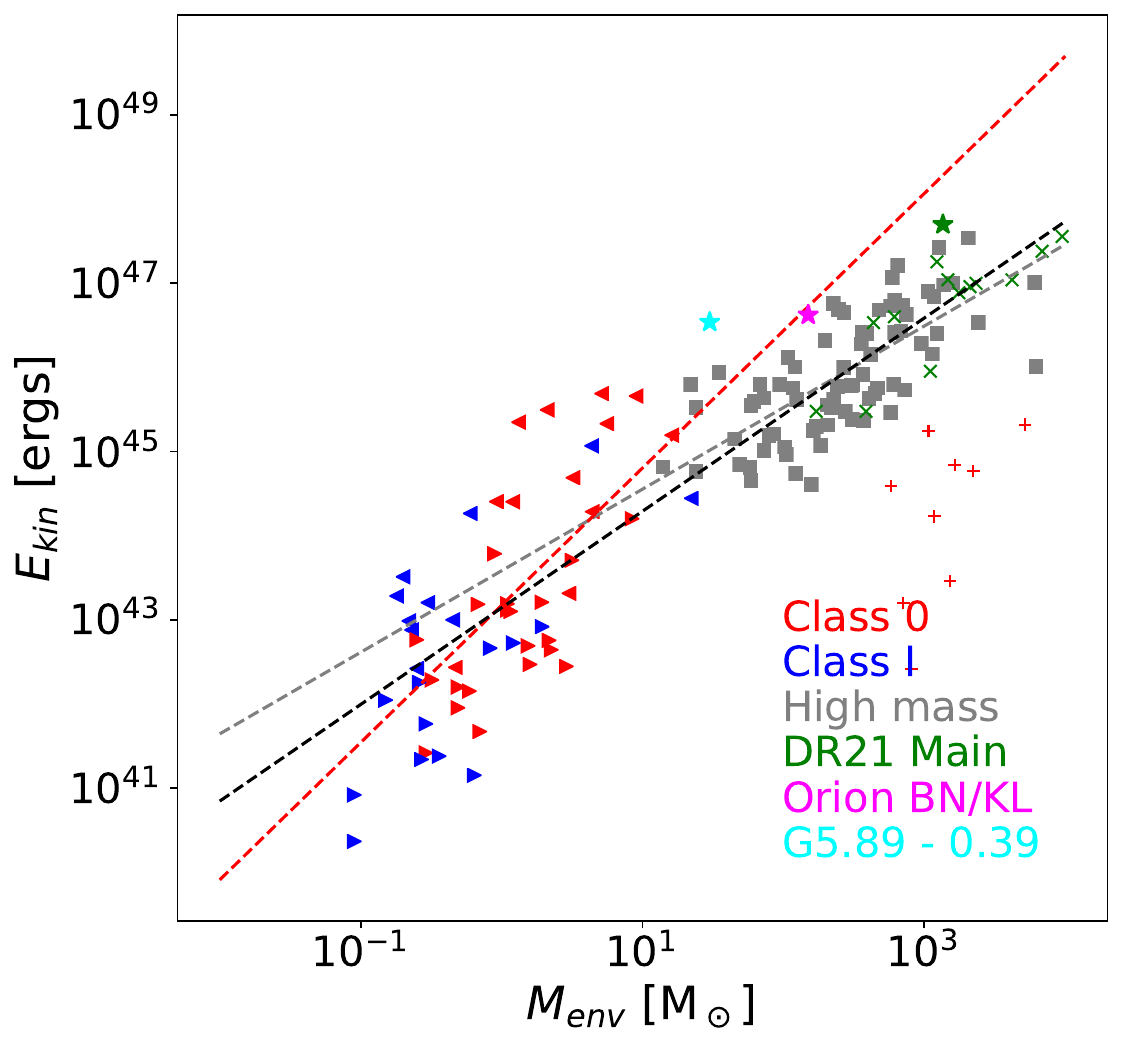}
    \includegraphics[width=8cm]{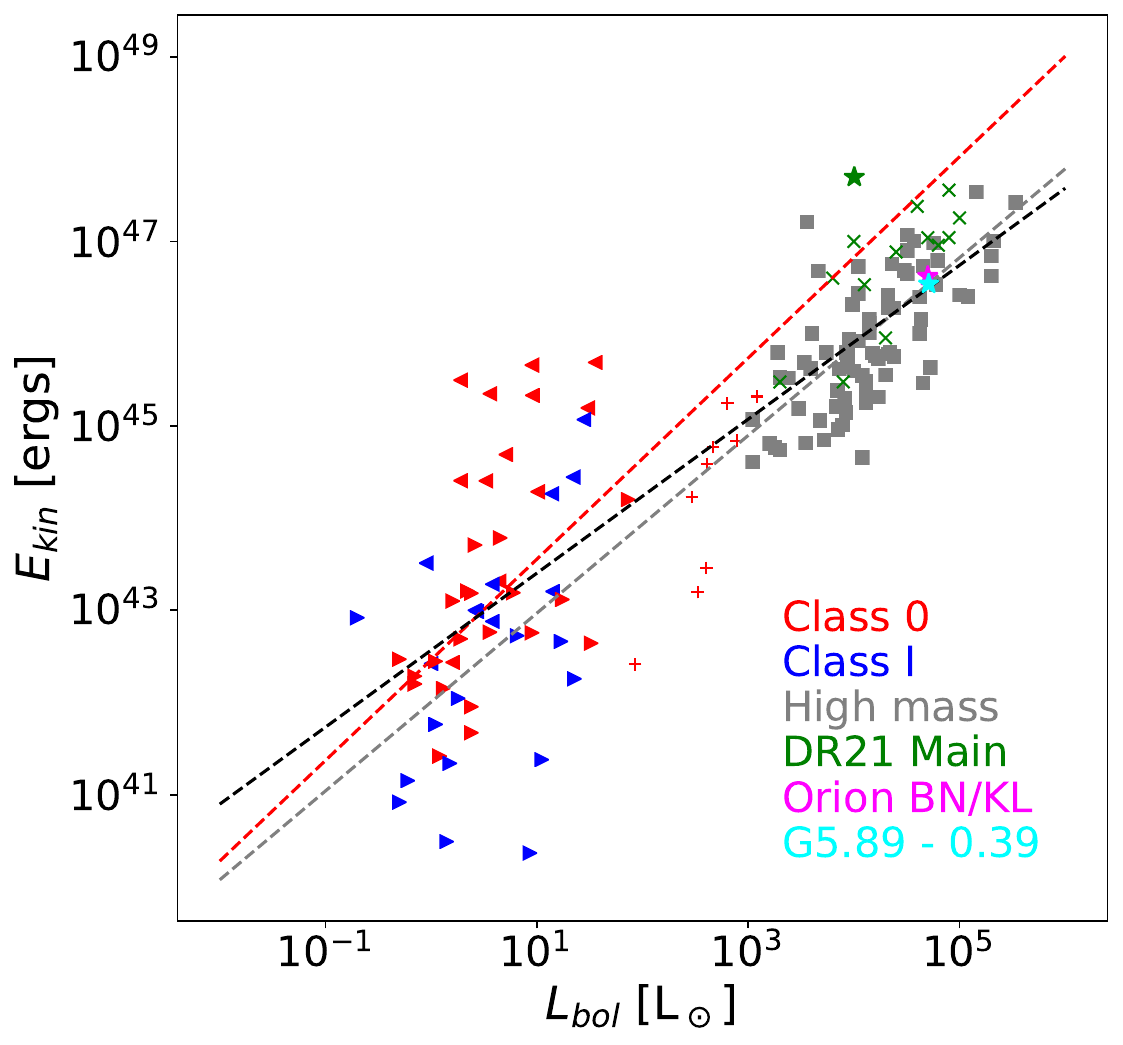}
    \caption{Outflow kinetic energy versus envelope mass and bolometric luminosity. The labels are the same as in Fig.~\ref{fig:forcemass}.}
    \label{fig:cor4}
\end{figure*}

\begin{figure*}
    \centering
    \includegraphics[width=8cm]{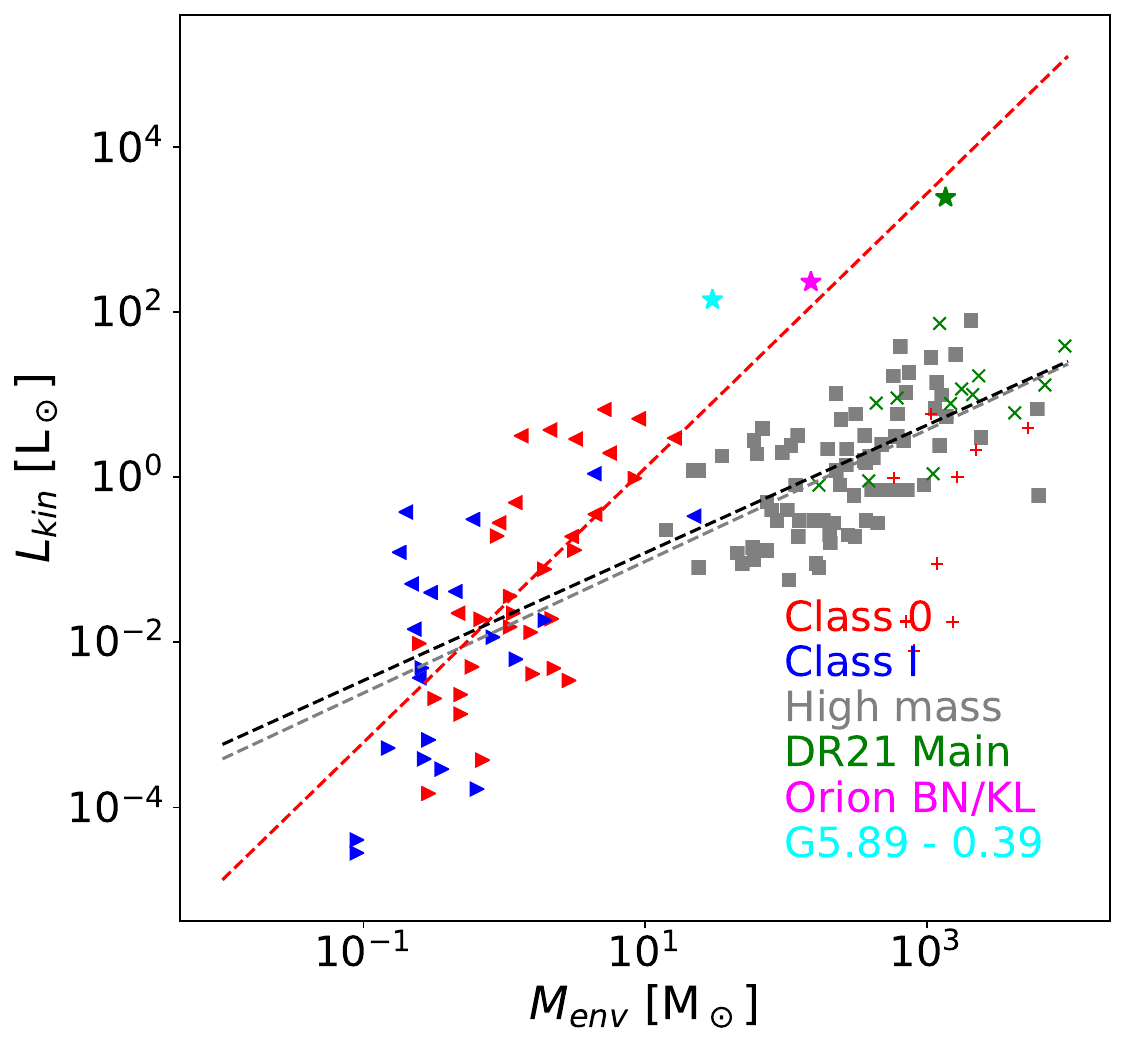}
    \includegraphics[width=8cm]{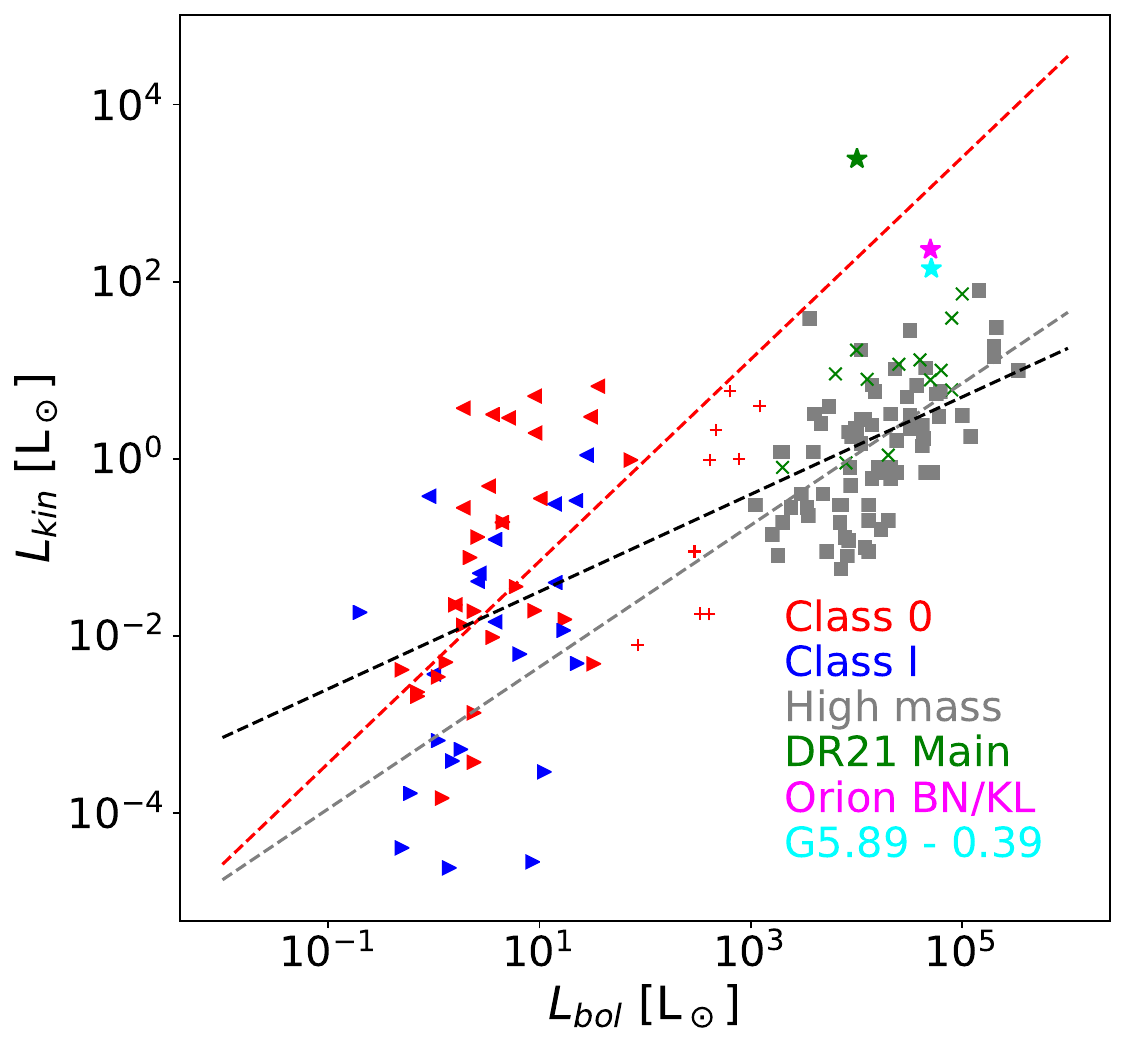}
    \caption{Outflow kinetic luminosity versus envelope mass and bolometric luminosity. The labels are the same as in Fig.~\ref{fig:forcemass}.}
    \label{fig:cor5}
\end{figure*}

Figures \ref{fig:cor1}--\ref{fig:cor5} show correlation plots between the different outflow parameters and the bolometric luminosity and envelope mass of their driving sources. 

\end{appendix}

\end{document}